\documentstyle[aps,psfig]{revtex}

\begin{document}

\title{Interplay of creation, propagation, and relaxation of an
excitation in a dimer}
\author{Jan Pe\v{r}ina, Jr.
\thanks{Present address: Joint Laboratory of Optics of Palack\'{y}
University and Institute of Physics of Academy of Sciences
of the Czech Republic, 17.~listopadu 50, 772 07 Olomouc,
Czech Republic} \\
Institute of Physics of Charles University,\\
Faculty of Mathematics and Physics,\\
Ke Karlovu 5, 121 16 Prague 2,\\
Czech Republic}

\date{}
\maketitle

\begin{abstract}
Interplay of simultaneous creation, annihilation,
propagation, and relaxation of an excitation in molecular
condensates interacting with an
ultrashort quantum optical pulse is studied
in general and specialized to a dimer.
A microscopic model appropriate for such systems
(with a strong exciton--phonon coupling) is presented.
It also incorporates effects of
(quantum) noise in the optical field.
A variety of new features in the initial stage of excitation dynamics
(when it is being created) is revealed;
a strong influence of the coherent excitation propagation
on the processes of excitation creation and annihilation
in a molecule strongly interacting with phonons is the most
remarkable one.
\end{abstract}



\vspace{4mm}
\noindent
{\Large \bf 1. Introduction}
\vspace{3mm}

Over the last decades, a great deal of attention has been
devoted to the study of
dynamics of excitations in systems with strong exciton--
or electron--phonon interactions like in molecular condensates
(\cite{ow1,ow2,ow3,ow4,ow5,ow6,ow7}; for review, see, Refs.
\cite{SilCa,Agra1,Agra2,Agra,Agra3}).
A strong exciton--phonon
interaction leads to polaron formation which results in
quasicoherent propagation of excitations. Quasicoherent
propagation has been extensively studied for various models
of the exciton--phonon interaction based on the (generalized)
master equation approach (for review, see, e.g., Ref.
\cite{SilCa}). However, all previously developed models assumed that
an excitation already exists at the beginning of the interation
with phonons (e.g., as a result of the interaction with
an ultrashort pulse).

An alternative approach to quasicoherent propagation
in molecular condensates is based
on the Green functions technique \cite{May,NMM}. But
it has in general a different range
of validity in comparison with the master
equation approach.

On the other hand, the optical Bloch equations provide
the description of a two-level system with relaxation interacting
with a classical optical field \cite{SM}.
This approach is appropriate in many experimental situations
(see, e.g., Refs. \cite{KSPSH,KSPSH1}). However, it fails
in systems interacting strongly with phonons. A generalization
of the Bloch equations in case of a two-level system interacting
strongly with phonons and a classical monochromatic optical field
has been found in Ref. \cite{GKS}. It has been also shown in Ref. \cite{GKS}
that field dependent relaxation rates emerge for strong optical
fields.
Also intermolecular coherent transfer influences
substantially excitation dynamics in systems interacting with
optical fields \cite{RKK,AS,KS,SAF,KF}.
Influence of statistical properties of optical fields in interaction
with simple matter systems (free and interacting atoms in resonant cavities,
atoms in traps, etc.) has been developed in quantum optics
(see, e.g., Refs. \cite{SM,AE,AE1,Per,JH1,JH2,KFL}).

However,
the above mentioned theories are not appropriate for
the description of excitation dynamics in molecular
condensates interacting with pulsed optical fields.
The paper provides such a theory, i.e. the theory applicable
to the description of excitation dynamics in systems
i) with coherent transfer of excitations, ii) with
a strong exciton--phonon interaction, and iii) being
under the influence of optical fields (e.g., in the
form of an ultrashort pulse).
It thus provides a tool for the investigation
of the initial stage of excitation dynamics,
i.e. when an excitation is being created.
Understanding of the excitation dynamics in the
initial stage is the main goal of the paper.
Moreover, the theory
encompasses also the influence of (quantum) noise in an
optical field.

The theory is based on projection operator
formalism \cite{pof1,pof2,pof3,pof4,pof5,pof6}.
The formalism of the theory is more complex in comparison with
the previously developed theories, but it is much
more general. In fact, it generalizes the theories of quasicoherent
transfer \cite{SilCa} which did not take
into account interaction with an optical field and thus
were not able to
describe at least the initial stage of excitation dynamics.
However, when the optical field is gone, the formalism
provides the same excitation dynamics as the older theories.

The general theory is specialized to a dimer interacting
with an ultrashort optical pulse.
The initial stage of excitation dynamics is then studied
in cases in which processes of excitation creation, transfer,
relaxation, and annihilation occur at the same timescale,
i.e. they mutually compete. In such cases, the previously derived
theories cannot be applied.
Influence of the coherent excitation
transfer on the processes of excitation creation and
annihilation in a molecule strongly interacting with
a phonon system is the most important effect.
Influence of noise in the optical field
is also investigated. It is shown, that the
theory provides the long-time excitation dynamics in agreement
with equilibrium statistical physics.

The theory is primarily developed for the description
of excitation dynamics in various molecular
condensates. However, it can also be applied to
other physical systems, e.g., to
the dynamics of an excitation in an impurity in a crystal.

Section 2 provides a general theory valid
for an arbitrary exciton system. It is specialized
to a dimer in Section 3. The initial stage of excitation
dynamics is investigated in Section 4.
Influence of an optical noise is studied in Section 5.
Section 6 is devoted to the long-time behaviour of an excitation.
Section 7 summarizes obtained results.
Appendix contains definitions of coefficients
entering equations for the dimer.

\vspace{4mm}
\noindent
{\Large \bf 2. Description of a general model}
\vspace{3mm}

A system under consideration appropriate for the description of excitation
dynamics in molecular condensates consists of the exciton (or electron),
photon, and phonon subsystems
with the exciton--photon and exciton--phonon interactions
(for details, see Refs. \cite{clanek1,diser}).
Hamiltonians of
the free exciton ($ \hat{H}_e $), photon ($ \hat{H}_f $), and phonon
($ \hat{H}_{ph} $) subsystems are given as follows
\begin{eqnarray}             % 1
 \hat{H}_e &=& \sum_{m,n} J_{mn} \hat{c}^\dagger_m \hat{c}_n , \nonumber \\
 \hat{H}_f &=& \sum_K \hbar\omega_{K} \hat{a}^\dagger_K \hat{a}_K ,
 \nonumber \\
 \hat{H}_{ph} &=& \sum_k \hbar\Omega_k \hat{b}^{\dagger}_k \hat{b}_k ,
\end{eqnarray}
where $ \hat{c}^\dagger_m $ ($ \hat{c}_m $) means the exciton creation
(annihilation) operator at the $ m $th site of the lattice
(Frenkel excitons are considered),
$ \hat{a}^\dagger_K $ ($ \hat{a}_K $)
represents the creation (annihilation) operator of the $ K $th photon
mode and $ \hat{b}^\dagger_k $ ($ \hat{b}_k $) is the creation
(annihilation) operator of the $ k $th phonon mode.
The exciton operators $ \hat{c}^\dagger_m $ and $ \hat{c}_m $ obey the Pauli
commutation relations \cite{Toyo}; i.e. the operator $ \hat{c}^\dagger_m $
can be expressed as $ \hat{c}^\dagger_m = \hat{d}^\dagger_m
\hat{e}_m $, where $ \hat{d}^\dagger_m $ is the creation operator
of an electron in the excited state and $ \hat{e}_m $ means
the annihilation operator of
an electron in the ground state at the $ m $th site.
The photon and phonon operators
$ \hat{a}^\dagger_K $, $ \hat{a}_K $, $ \hat{b}^\dagger_k $, and $ \hat{b}_k $
obey the boson commutation relations.
The coefficients
$ J_{mn} $ describe energies of the free exciton subsystem for $ m=n $
and coherent transfer in the exciton subsystem given by overlaps of
wave functions
for $ m \neq n $. The excitonless state $ |0\rangle $ is
assumed to have zero energy. The symbol $ \omega_K $ ($ \Omega_k $) stands
for the frequency of the $ K $th ($ k $th) mode of the photon (phonon) field.
The symbol $ \sum_K $ ($ \sum_k $) means summation over all photon (phonon)
modes and $ \sum_m $ denotes summation over all exciton states; $ \hbar $ is
the reduced Planck constant.

The exciton--photon interaction Hamiltonian $ \hat{H}_{e-f} $ in the rotating
wave approximation reads \cite{Toyo}
\begin{equation}             % 2
 \hat{H}_{e-f} = \sum_{m,K} \hbar\omega_{K_0} F^m_K \left( \hat{a}_K
 \hat{c}^\dagger_m + \hat{a}^\dagger_{-K} \hat{c}_m \right) ,
\end{equation}
where $ \omega_{K_0} $ is a typical photon frequency.
The exciton--photon coupling constants $ F^m_K $ are given in the dipole
approximation by
\begin{equation}             % 3
 F^m_K = - \frac{1}{\hbar\omega_{K_0}}
\sqrt{\frac{\hbar}{2\epsilon_0
 V\omega_K}}
 \frac{e}{m_e} \mbox{\boldmath $ \epsilon $}_K \cdot \langle m|\hat{\bf p}
 |0\rangle \exp \left( i{\bf K} \cdot {\bf r}_m \right) .
\end{equation}
Here $ e $ ($ e<0 $) is the charge of electron, $ m_e $ the mass of electron,
$ \hat{\bf p} $ the momentum operator of electron, $ {\bf r}_m $
the mean position
of the $ m $th molecule in a lattice, $ |0\rangle $ describes
the excitonless
state, $ |m\rangle $ the state with one exciton localized at the $ m $th
site of the lattice
and the dot means the scalar product; $ \mbox{\boldmath $ \epsilon $}_K $ is
the polarization vector of the $ K $th mode of the photon field,
$ \epsilon_0 $ permittivity of vacuum, and $ V $
the quantization volume of the electromagnetic field.

The exciton--phonon interaction is described by the interaction Hamiltonian
$ \hat{H}_{e-ph} $ in the form \cite{Toyo,SilCa}
\begin{equation}             % 4
 \hat{H}_{e-ph} = \frac{1}{\sqrt{N}} \sum_{m,k} \hbar\Omega_k G^m_k
 \hat{c}^\dagger_m \hat{c}_m \left( \hat{b}_k + \hat{b}^\dagger_{-k} \right)
\end{equation}
with only the site-diagonal exciton--phonon coupling included.
The dependence of the exciton--phonon
coupling constants $ G^m_k $ on the site index $ m $ and the mode index $ k $
is determined according to the type of phonons (optical or acoustic)
and according to the model of the exciton--phonon interaction; $ N $ means
the number of phonon modes. Hamiltonian (4) describes a deformation of the
lattice
around a given site $ m $ after it was occupied by the exciton
(polaron effect).

The excitation dynamics in such a complex system can
be conveniently described by the generalized master equations
for the exciton reduced density matrix $ \hat{\rho}(t) $.
The application of the time-convolutionless formalism
\cite{FC} in
connection with a time-dependent projector \cite{TP1}
and the assumption that $ \hat{\rho}(t) $ lies within the
space corresponding to the Hilbert space spanned by the
excitonless state $ |0\rangle $ and by states
$ |m\rangle $ with one
exciton at a given site $ m $
(``single-excitation approximation'')
result in the following set of equations for the matrix
elements of the exciton reduced density matrix
$ \hat{\rho}(t) $
(for details, see Refs. \cite{clanek1,diser});
\begin{eqnarray}             % 5-7
 \frac{d}{dt}\rho_{mn}(t) &=& - \frac{i}{\hbar} \sum_p J_{mp}\rho_{pn}(t)
  + \frac{i}{\hbar} \sum_p J_{pn}\rho_{mp}(t)
  - \sum_p G_{mn,p}(t)\rho_{pn}(t) - \sum_p G^*_{nm,p}(t) \rho_{mp}(t)
  \nonumber \\
 & & \mbox{} - \sum_p \bar{I}_{m,p}(t)\rho_{pn}(t) - \sum_p
 \bar{I}^*_{n,p}(t)
  \rho_{mp}(t) \nonumber \\
 & & \mbox{} - i F_m(t)\rho_{0n}(t) + i F^*_n(t) \rho_{m0}(t)
  + \left[ I_{m,n}(t) + I^*_{n,m}(t) \right] \rho_{00}(t) , \\
 \frac{d}{dt} \rho_{00}(t) &=& - \sum_l \left[ I^*_{l,l}(t)
  + I_{l,l}(t) \right]  \rho_{00}(t) \nonumber \\
 & & \mbox{} - i \sum_p F^*_{p}(t) \rho_{p0}(t)
  + i \sum_p F_p(t) \rho_{0p}(t)
  + \sum_{s,p} \left[ \bar{I}^*_{s,p}(t) + \bar{I}_{p,s}(t) \right]
  \rho_{sp}(t) , \\
 \frac{d}{dt}\rho_{0n}(t) &=& \frac{i}{\hbar} \sum_p J_{pn}\rho_{0p}(t)
  - \sum_p G^*_{n,p}(t)\rho_{0p}(t) - \sum_{p} I^*_{n,p}(t)
  \rho_{0p}(t) - \sum_l I^*_{l,l} (t) \rho_{0n}(t) \nonumber \\
 & & \mbox{} + \sum_p \left[ \tilde{I}_{p,n}(t) + \tilde{I}_{n,p}(t) \right]
  \rho_{p0}(t) - i \sum_p F^*_{p}(t)\rho_{pn}(t) + i F^*_{n}(t)
  \rho_{00}(t) .
\end{eqnarray}
The matrix element $ \rho_{mm} $ describes the
probability that an exciton is in a state localized at the site $ m $
and $ \rho_{00} $
determines the probability of the system to be in the
excitonless state. The nondiagonal elements $ \rho_{mn} $ ($ m \ne n $)
containing information about the mutual coherence between states
localized at sites $ m $ and $ n $ play the dominant role in the
description of quasicoherent excitation propagation.
The elements $ \rho_{0m} $ reflecting a mutual coherence
between the excitonless state and excited states
are important for the description of
excitation creation and annihilation.

The coefficients $ J_{mn} $ describe the inner dynamics of the
exciton subsystem (coherent transfer).
The time-dependent coefficients $ G_{mn,p}(t) $ and $ G_{m,p}(t) $
originating in the exciton--phonon interaction describe effects
of polaron formation. They renormalize the coefficients
$ J_{mn} $ (see Section 6 later). This results in quasicoherent
excitation propagation. The time-dependent coefficients
$ F_m(t) $ are responsible for excitation creation and
annihilation caused by the coherent part of the optical field,
whereas the coefficients $ I_{m,p}(t) $, $ \bar{I}_{m,p}(t) $, and
$ \tilde{I}_{m,p}(t) $ reflect effects of the incoherent
part of the optical field (noise).

The time-dependent coefficients in Eqs. (5---7)
are correct to the second power of the exciton--photon or
exciton--phonon coupling constants. However, this does not
mean that the solution of Eqs. (5---7) is also restricted
to the second order.

The time-dependent coefficients $ G_{mn,p}(t) $ and $ G_{m,p}(t) $
are given as follows
\begin{eqnarray}             % 8
 G_{mn,p}(t) &=& \int^{t-t_0}_{0} d\tau \, \frac{1}{N} \sum_k \Omega^2_k
  \left( G^m_k - G^n_k \right)
  \left\{ \left[ n_{\rm B}(\hbar\Omega_k) + 1 \right] \exp\left( -i\Omega_k
  \tau\right) +  n_{\rm B}(\hbar\Omega_{-k}) \exp\left(i\Omega_{-k}
  \tau  \right)
  \right\} \nonumber \\
 & & \mbox{} \times \sum_s G^s_{-k} \langle m|s\rangle(\tau) \langle p|s
  \rangle^*(\tau) , \nonumber \\
 G_{n,p}(t) &=& \int^{t-t_0}_{0} d\tau \, \frac{1}{N} \sum_k \Omega^2_k G^n_k
  \left\{ \left[ n_{\rm B}(\hbar\Omega_k) + 1 \right] \exp\left( -i\Omega_k
  \tau\right) +  n_{\rm B}(\hbar\Omega_{-k}) \exp\left(i\Omega_{-k}
  \tau \right)
  \right\} \nonumber \\
 & & \mbox{} \times \sum_s G^s_{-k} \langle n|s\rangle(\tau) \langle
  p|s\rangle^*(\tau) ,
\end{eqnarray}
where the symbol
\begin{equation}             % 9
 n_{\rm B}(\hbar\Omega_k) = \frac{1}{ \exp(\beta\hbar\Omega_k) -1}
\end{equation}
denotes the mean value of the number of phonons in the mode $ K $ being in
the equilibrium state and the correlation function $ \langle p|s\rangle(t) $
of the noninteracting exciton subsystem is defined by
\begin{equation}             % 10
 \langle p|s\rangle(t) = \langle p| \exp \left( - \frac{i}{\hbar} \hat{H}_e t
 \right) |s\rangle .
\end{equation}

The time-dependent coefficients
$ F_m(t) $, $ I_{m,p}(t) $, $ \bar{I}_{m,p}(t) $, and
$ \tilde{I}_{m,p}(t) $
are expressed in the form
\begin{eqnarray}             % 11
 F_m(t) &=& \omega_{K_0} \tilde{F}^m_{K_0} {\cal A}(t) , \nonumber \\
 I_{m,p}(t) &=& \int^t_{t_0} d\tau \, \omega^2_{K_0} \tilde{F}^m_{K_0} \sum_s
  \tilde{F}^s_{-K_0} \delta N(t,\tau) \langle p|s\rangle^*(t-\tau) ,
  \nonumber \\
 \bar{I}_{m,p}(t) &=& \int^t_{t_0} d\tau \, \omega^2_{K_0}\tilde{F}^m_{K_0}
  \sum_s \tilde{F}^s_{-K_0} \delta N_v(t,\tau) \langle p|s\rangle^*(t-\tau) ,
  \nonumber \\
 \tilde{I}_{m,p}(t) &=& \int^t_{t_0} d\tau \, \omega^2_{K_0}\tilde{F}^m_{-K_0}
  \sum_s \tilde{F}^s_{-K_0} \delta N_a(t,\tau) \langle
  p|s\rangle^*(t-\tau),
\end{eqnarray}
where
\begin{equation}             % 12
 \tilde{F}^m_{K_0} = - \frac{1}{\hbar\omega_{K_0}} \frac{e}{m_e}
  \mbox{\boldmath $ \epsilon $}_{K_0} \cdot \langle m|\hat{\bf p}
  |0\rangle \exp \left( i{\bf K}_0 \cdot {\bf r}_m \right).
\end{equation}
The optical field is assumed to be polarized in the
direction $ \mbox{\boldmath $ \epsilon $}_{K_0} $.

The function
\begin{equation}             % 13
 {\cal A}(t) = \sum_K \sqrt{ \frac{\hbar}{2\epsilon_0 V\omega_K} } \langle
 \alpha_K(t)\rangle_f
\end{equation}
describes a classical amplitude of the field. Second moments
of the optical field are
characterized by the functions
\begin{eqnarray}             % 14
 \delta N(t,\tau) &=& \sum_{K,K'} \frac{\hbar}{2\epsilon_0 V \sqrt{
  \omega_K\omega_{K'}} } \langle \delta\alpha_K(t) \delta\alpha^*_{K'}(\tau)
  \rangle_f , \nonumber \\
 \delta N_v(t,\tau) &=& \delta N(t,\tau) + \sum_K \frac{\hbar}{
  2\epsilon_0 V \omega_K} \exp \left[-i\omega_K(t-\tau) \right] , \nonumber \\
 \delta N_a(t,\tau) &=& \sum_{K,K'} \frac{\hbar}{2\epsilon_0V \sqrt{
  \omega_K\omega_{K'}} } \langle \delta\alpha^*_K(t) \delta\alpha^*_{K'}(\tau)
  \rangle_f .
\end{eqnarray}
The above introduced correlation functions of the photon field
operators are defined as follows:
\begin{eqnarray}             % 15
 \langle \alpha_K(t) \rangle_f &=& {\rm Tr }_f \left\{ \hat{\rho}_f(t_0)
  \hat{a}_K(t-t_0) \right\} , \nonumber \\
 \langle \alpha_K(t) \alpha^*_{K'}(\tau) \rangle_f &=& {\rm Tr }_f \left\{
  \hat{\rho}_f(t_0)
  \hat{a}^\dagger_{K'}(\tau-t_0) \hat{a}_K(t-t_0) \right\} , \nonumber \\
 \langle \alpha^*_K(t) \alpha^*_{K'}(\tau) \rangle_f &=& {\rm Tr }_f
  \left\{ \hat{\rho}_f(t_0)
  \hat{a}^\dagger_{K}(t-t_0) \hat{a}^\dagger_{K'}(\tau-t_0) \right\},
  \nonumber \\
 \delta\alpha_K(t) &=& \alpha_K(t) - \langle \alpha_K(t)\rangle_f .
\end{eqnarray}

A detailed analysis of the above general equations is
contained in Refs. \cite{clanek1,diser}.

\vspace{4mm}
\noindent
{\Large \bf 3. Asymmetric dimer}
\vspace{3mm}

The model developed in the previous section is now
specialized to a dimer consisting in general of
two different molecules.
We assume that the first (second) molecule has the energy
$ E+2\varepsilon $ ($ E $) and that coherent
exciton transfer between molecules is described by
a real constant $ J $. Hence
\begin{equation}             % 16
 J_{11} = E+2\varepsilon , \hspace{1cm}
 J_{22} = E , \hspace{1cm}
 J_{12} = J_{21} = J .
\end{equation}
The exciton Hamiltonian (16) can be easily diagonalized by
the eigenvectors
\begin{eqnarray}             % 17
 |+\rangle &=& \frac{1}{ \sqrt{ 2\Delta (\Delta - \varepsilon)} } \left[ J
 |1\rangle + ( -\varepsilon +\Delta ) |2\rangle \right] , \nonumber \\
 |-\rangle &=& \frac{1}{ \sqrt{ 2\Delta (\Delta + \varepsilon)} } \left[ J
 |1\rangle + ( -\varepsilon -\Delta) |2\rangle \right]
\end{eqnarray}
with the corresponding eigenenergies
\begin{eqnarray}             % 18
 E_+ &=& E + \varepsilon + \Delta , \nonumber \\
 E_- &=& E + \varepsilon - \Delta ,
\end{eqnarray}
where
\begin{equation}             % 19
 \Delta = \frac{1}{2} (E_+ - E_-) = \sqrt{ \varepsilon^2 + J^2} .
\end{equation}

This enables us to calculate the correlation functions
$ \langle p|s\rangle(t) $ in Eq. (10)
and then to determine the time-dependent coefficients entering
Eqs. (5---7).
The resulting equations
represent a set of nine coupled differential equations with
a special structure. They can be conveniently written in the
matrix form:
\begin{equation}             % 20
 \frac{d}{dt} \left[ \begin{array}{c} {\cal R}_1(t) \\ {\cal R}_2(t)
  \end{array} \right]
  = \left[ \begin{array}{cc} {\cal J}_1 & 0 \\ 0 & {\cal J}_2 \end{array}
  \right]
  \left[ \begin{array}{c} {\cal R}_1(t) \\ {\cal R}_2(t) \end{array} \right]
  - \left[ \begin{array}{cc} {\cal G}_1(t) & 0 \\ 0 & {\cal G}_2(t)
  \end{array} \right]
   \left[ \begin{array}{c} {\cal R}_1(t) \\ {\cal R}_2(t) \end{array} \right]
  - \left[ \begin{array}{cc} {\cal F}_1(t) & {\cal F}_2(t) \\ {\cal F}_3(t) &
  {\cal F}_4(t) \end{array} \right]
   \left[ \begin{array}{c} {\cal R}_1(t) \\ {\cal R}_2(t) \end{array}
   \right] .
\end{equation}
The vectors $ {\cal R}_1(t) $ and $ {\cal R}_2(t) $ are defined
as follows;
\begin{eqnarray}             % 21, 22
 {\cal R}_1(t) &=& \left[  \begin{array}{c} \rho_{11}(t) \\
  \rho_{22}(t) \\
  \rho_{r}(t) \\ \rho_{i}(t) \\ \rho_{00}(t) \end{array} \right] , \\
 {\cal R}_2(t) &=& \left[  \begin{array}{c} \rho_{1r}(t) \\
  \rho_{1i}(t) \\
  \rho_{2r}(t) \\ \rho_{2i}(t) \end{array} \right] .
\end{eqnarray}

The exciton matrix elements in Eqs. (21) and (22)
are expressed as follows:
\begin{eqnarray}             % 23
 \rho_r(t) &=& \frac{1}{2} \left[ \rho_{12}(t) + \mbox{c.c.} \right] ,
  \nonumber \\
 \rho_i(t) &=& \frac{1}{2i} \left[ \rho_{12}(t) - \mbox{c.c.} \right] ,
\end{eqnarray}
\begin{eqnarray}             % 24
 \rho_{jr}(t) &=& \frac{1}{2} \left\{ \rho_{0j}(t) \exp\left[
  -\frac{i}{\hbar}(E+\varepsilon)t \right] + \mbox{c.c.} \right\} ,
  \nonumber \\
 \rho_{ji}(t) &=& \frac{1}{2i} \left\{ \rho_{0j}(t) \exp\left[
  -\frac{i}{\hbar}(E+\varepsilon)t \right] - \mbox{c.c.} \right\} ,
   \hspace{1cm} j=1,2 ,
\end{eqnarray}
where $ \mbox{c.c.} $ means complex conjugate.

The matrices $ {\cal J}_1 $ and $ {\cal J}_2 $ describe coherent
transfer in the free exciton subsystem.
Effects of polaron formation as a result of the exciton--phonon interaction
are contained in the matrices $ {\cal G}_1(t) $ and
$ {\cal G}_2(t) $.
Excitation creation and annihilation caused by the
coherent part of a
photon field is described in the matrices $ {\cal F}_2(t)
$ and $ {\cal F}_3(t) $. Finally, the matrices $ {\cal F}_1(t) $ and
$ {\cal F}_4(t) $ contain the influence
of noise in the photon field.
Definitions of these matrices
as well as the description
of their role in excitation dynamics
are contained in Appendix. We limit ourselves only to
the discussion of main characteristic features here.

The matrices on the right-hand side of Eq. (20) have a remarkable block
structure. The dynamics of the free
exciton subsystem is governed by the
mutually independent vectors
$ {\cal R}_1(t) $ and $ {\cal R}_2(t) $ as a result
of the special type of the exciton Hamiltonian in Eq. (1).
Owing to the number of excitation conservation as a consequence
of our special form of the exciton--phonon coupling (4),
the latter interaction changes the dynamics of $ {\cal R}_1(t) $ and
$ {\cal R}_2(t) $, but it does not change their mutual independence.
Noise in the photon field acts similarly.
The coherent part of an optical field
introduces mutual coupling between the vectors
$ {\cal R}_1(t) $ and $ {\cal R}_2(t) $
and thus leads to an effective exciton generation.
We note that exciton generation can also be caused
by noise in the optical field (for details, see Appendix).

Interaction with the coherent component of the photon field
creates two different kinds of paths leading to excitation
generation.
In the first kind, for molecule~1, an excitation (nonzero $ \rho_{11}(t) $)
emerges along the path $ \rho_{00} \stackrel{F^1_{K_0}}{\rightarrow}
\rho_{10}, \rho_{01} \stackrel{F^1_{K_0}}{\rightarrow} \rho_{11} $
including double interaction of the photon field at this
molecule. Terms corresponding to such a path are the same
as those in the exact equations for a two-level system
interacting with a classical deterministic time-dependent
field \cite{diser}. In this case, the solution
of our equations is nonperturbative with respect to
the exciton--photon coupling constants.
For the second kind of excitation paths, the
existence of a surrounding molecule is necessary. An excitation
can emerge along the paths $ \rho_{00} \stackrel{F^1_{K_0}}{\rightarrow}
\rho_{10}, \rho_{01} \stackrel{F^2_{K_0}}{\rightarrow} \rho_{12},
\rho_{21} \stackrel{J}{\rightarrow} \rho_{11} $ and $ \rho_{00}
\stackrel{F^2_{K_0}}{\rightarrow} \rho_{20}, \rho_{02}
\stackrel{F^1_{K_0}}{\rightarrow} \rho_{12}, \rho_{21}
\stackrel{J}{\rightarrow} \rho_{11} $ including double interaction with
the photon field at molecules 1 and 2 and transfer through $ J $.

The statistics of an optical field is not
limited to a classical noise. Quantum description
of the optical field provides a tool for investigations of
the influence of nonclassical properties of light
(e.g. squeezing of vacuum fluctuations).

Time development of the coefficients originating in
exciton--phonon coupling reflects
a polaron cloud formation around the exciton.
The coefficients are practically zero for short
times and start to act significantly for longer times when
they renormalize coefficients in the matrices
$ {\cal J}_1 $ and $ {\cal J}_2 $.
Their time development ceases for
times comparable with the phonon relaxation time $ \tau_R $ when they reach
asymptotic values. The magnitude of renormalization is proportional to the
strength of exciton--phonon coupling. Asymptotic values of
the renormalization are given by the Debye-Waller factor as will be
shown in Section 6. That means that our model, although
perturbative in exciton--phonon coupling, is able to
describe correctly effects of a strong exciton--phonon coupling
(small polaron formation).

The above presented equations for a dimer (20) extend Bloch equations
in three respects: i) they include transfer of an excitation from and
to a given molecule, ii) they describe interaction with
a phonon reservoir on a ``microscopic level'' (polaron
formation), and finally iii) they incorporate effects
of (quantum) noise in the optical field.

\vspace{4mm}
\noindent
{\Large \bf 4. Initial stage of excitation dynamics}
\vspace{3mm}

Based on the above introduced model of the dimer, excitation dynamics is
investigated
under the conditions when times characterizing pulse
duration, propagation, and relaxation are comparable.
In this case, the model provides new interesting results.
Firstly, excitation dynamics is studied in a two-level system
strongly interacting with phonons (Subsection 4.1). In comparison with
Ref. \cite{GKS} the model is valid also for pulsed light
and provides a more accurate description
of effects stemming from the exciton--phonon interaction.
Effects of excitation transfer are studied in a dimer
in which either one molecule interacts with a
pulse (Subsection 4.2) or both molecules interact
with a pulse simultaneously (Subsection 4.3).
Such models are appropriate for a variety
of physical situations (e.g. two interacting
molecules of the same or different kind, a molecule surrounded
by an environment with which it can exchange energy,
etc.).

Investigation of the excitation dynamics is based on
the numerical solution of Eq. (20) for the model of the
exciton--phonon interaction discussed in Appendix
(for details, see, Ref. \cite{diser}).
The real envelope of the optical field
$ \tilde{\cal A}(t) $ (for definition, see (A9) in Appendix)
is assumed in the form
$ \tilde{\cal A}(t) = {\bf A} \tilde{\cal A}_n(t) $, where $ {\bf A} $
determines the strength of the optical field and
the function $ \tilde{\cal A}_n(t) $ has the form
\begin{eqnarray}             % 25
 \tilde{\cal A}_n(t) &=& 1 , \hspace{3.8cm} \mbox{for $ t\leq \tau_1 $}
 \nonumber  \\
 &=& \exp\left( -\frac{t-\tau_1}{\tau_2} \right) , \hspace{1.5cm}
  \mbox{for $ t>\tau_1 $} .
\end{eqnarray}
Pulse duration is characterized by the constant
$ \tau_1+\tau_2 $. The constants
$ F_1 = \hbar\omega_{K_0}\tilde{F}^1_{K_0}{\bf A} $
and $ F_2 = \hbar\omega_{K_0}\tilde{F}^2_{K_0}{\bf A} $
are used in the further discussion.

We introduce the following simplified notation:
$ p_0(t) $, $ p_1(t) $, and $ p_2(t) $ denote the
probabilities that
the exciton system is in the excitonless state ($ p_0(t) = \rho_{00}(t) $),
in the state localized at molecule 1 ($ p_1(t) = \rho_{11}(t) $), and
in the state localized at molecule
2 ($ p_2(t) = \rho_{22}(t) $); $ \rho_r(t) $ and $ \rho_i(t) $
mean the real and imaginary parts of $ \rho_{12}(t) $.
Meaning of parameters of the
system under consideration is schematically shown in Fig.~1.
The energy and time scales are introduced so that quantities in
energy units are in eV and time is in femtoseconds.
\begin{figure}        % fig. 1
  \centerline{\hbox{\psfig{file=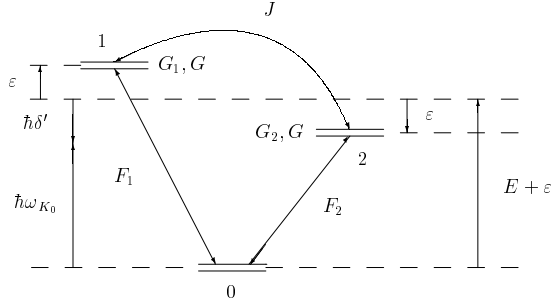,height=4cm} }}
  \vspace{3mm}
  \caption{The scheme of asymmetric dimer; $ F_1 $ and $ F_2 $ are
 the exciton--photon coupling constants at molecules 1 and 2, $ G_1 $ and
 $ G_2 $ the exciton--phonon coupling constants at molecules 1 and 2,
 $ G $ is the mean exciton--phonon coupling constant,
 $ E + \varepsilon $ the mean difference of energies between
 the excitonless state and
 the excited states, $ 2\varepsilon $ the energy difference
 between the excited states at molecules 1 and 2,
 $ J $ the transfer integral,
 $ \omega_{K_0} $ the carrying frequency of the optical field
 and $ \delta' $ describes the detuning of the optical field.
 The splitting ($ 2\varepsilon $) of excited levels can originate in
 different kinds of molecules or in the influence of
 local environment of molecules.}
\end{figure}

\vspace{3mm}
\noindent
{\it 4.1. Two-level system}
\vspace{2mm}

A two-level system interacting with an
optical field
\footnote{
 The magnitude of exciton--photon
 coupling constants in energy
 units can be estimated from the mean value of the exciton--photon
 interaction
 Hamiltonian $ \hat{H}_{e-f} $ as follows: $ \langle \hat{H}_{e-f}\rangle
 \approx
 \frac{e}{m_e} \langle\hat{p}\rangle A \approx \frac{e}{\hbar}\langle
 [\hat{H}_{e},\hat{x}]\rangle
 \frac{E}{\omega_0} \approx \frac{e}{\hbar} ({\cal E}_e - {\cal E}_g)
 \langle \hat{x}\rangle \frac{E}{\omega_0} \approx \frac{e}{\hbar}
 ({\cal E}_e -
 {\cal E}_g) a_{\rm B} \frac{E}{\omega_0} $, where $ A $ is the vector
 potential of a classical field, $ E $ the intensity of the
 field, $ \omega_0 $ the mean frequency of the field,
 $ \hat{p} $ ($ \hat{x} $) means the momentum (position) operator of electron,
 $ {\cal E}_e $ ($ {\cal E}_g $) denotes
 the energy of the excited
 (ground) state, and $ a_{\rm B} $ is the Bohr radius; $ E = 10^7
 \mbox{ Vm}^{-1} $,
 $ \lambda_0 = 600 \mbox{ nm} $, and $ {\cal E}_e - {\cal E}_g = 2 $~eV
 implies $ \langle\hat{H}_{e-f}\rangle \approx 5.10^{-4} \mbox{ eV}
 $.}
shows Rabi oscillations \cite{SM} (the stronger the field the
higher the frequency is).
Detuning ($ \delta' $) between
the carrying frequency of the
pulse and the frequency of the two-level system means faster
oscillations
(with the generalized Rabi frequency) and a lower excitation level
(smaller $ p_1(t) $).
When the strength of the optical field is
constant the system can return to the excitonless state,
but the excitonless state cannot be reached
in the period of the pulse quenching.

Interaction with phonons leads to polaron formation around
the exciton. This causes a successive diminishing of
the effective strength of exciton--photon coupling and thus
smaller values of the probability $ p_1(t) $
(it cannot take on value 1 for a nonzero
exciton--phonon coupling constant $ G $) (see Fig.~2).
This diminishing originates in
lower effective values of the electric dipole moment
$ -e\langle \hat{\bf x}\rangle $ caused by
a successive formation of the polaron state from a bare excited
state and in renormalization of the two-level system energy
leading to a greater effective detuning $ \delta' $.
Strong exciton--phonon coupling suppresses
deexcitation of the two-level system (see curves B, C, and D in
Fig.~2). Effects of polaron formation are more pronounced
for higher temperatures,
i.e. for greater mean numbers of equilibrium phonons $ n_{\rm B} $.
\begin{figure}        % fig. 2
  \centerline{\hbox{\psfig{file=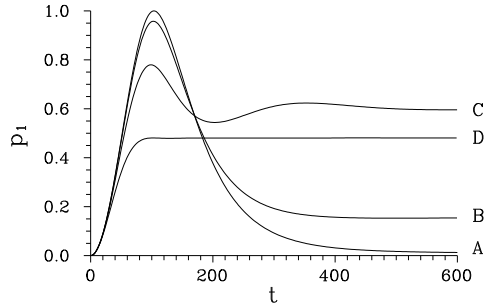,height=4cm} }}
  \vspace{3mm}
  \caption{The influence of the exciton--phonon coupling
  constant $ G $ on the probability $ p_1(t) $;
  $ G = 0 $ [A], $ G = 0.004 $ [B],
  $ G = 0.01 $ [C], $ G = 0.02 $ [D];
  $ J = 10^{-8} $, $ \hbar\delta' = 0 $,
  $ F_1 = 0.01 $, $ F_2 = 0 $,
  $ \varepsilon = 0 $, $ \tau_1 = 100 $, $ \tau_2 = 100 $,
  $ G_1/G = 1+0.25i $, $ G_2/G = 1-0.25i $,
  $ n_{\rm B} = 0 $, $ \hbar\Omega_{ph} = 0.01 $,
  $ \hbar\gamma_{ph} = 0.001 $.}
\end{figure}

\newpage
\vspace{3mm}
\noindent
{\it 4.2. Dimer with one molecule interacting with a pulse}
\vspace{2mm}

Only molecule 1 is assumed to interact with the pulse as, e.g.,
a consequence of a special orientation of the transition dipole
momentum of molecule 2.
In case of exciton dynamics, the ``single-excitation''
approximation requires that values of the whole excitation
probability of the dimer $ p_1(t) + p_2(t) $ are approximately
less than 0.6 (for details, see Subsection 4.3).
The obtained results are appropriate also for electron transfer
as a result of the ``single-excitation'' approximation which
rules out differencies originating in Fermi and Pauli
commutation relations.
In this case, the above limitation does not apply.

\vspace{3mm}
\noindent
{\it 4.2.1. Energetically balanced dimer}
\vspace{2mm}

The effect of {\it coherent transfer} (described by $ J $)
on the dynamics of the dimer
noninteracting with phonons is shown in Figs. 3a,b. The increase of
$ J $ leads to a faster exchange of excitation between molecules
1 ($ p_1(t) $) and 2 ($ p_2(t) $)
(its frequency is given by $ J $). The transfer
strongly affects the processes of excitation creation
and annihilation at molecule 1.
Fig.~3a shows that the value of $ J $ affects
the time in which the excitation of molecule 1 is replaced by its
deexcitation;
especially greater values of $ J $ mean an earlier time of deexcitation.
This results in smaller values of the whole excitation
probability $ p_1(t)+p_2(t) $ for greater values of $ J $.
\begin{figure}        % figs. 3a,b
  \centerline{\hbox{ \psfig{file=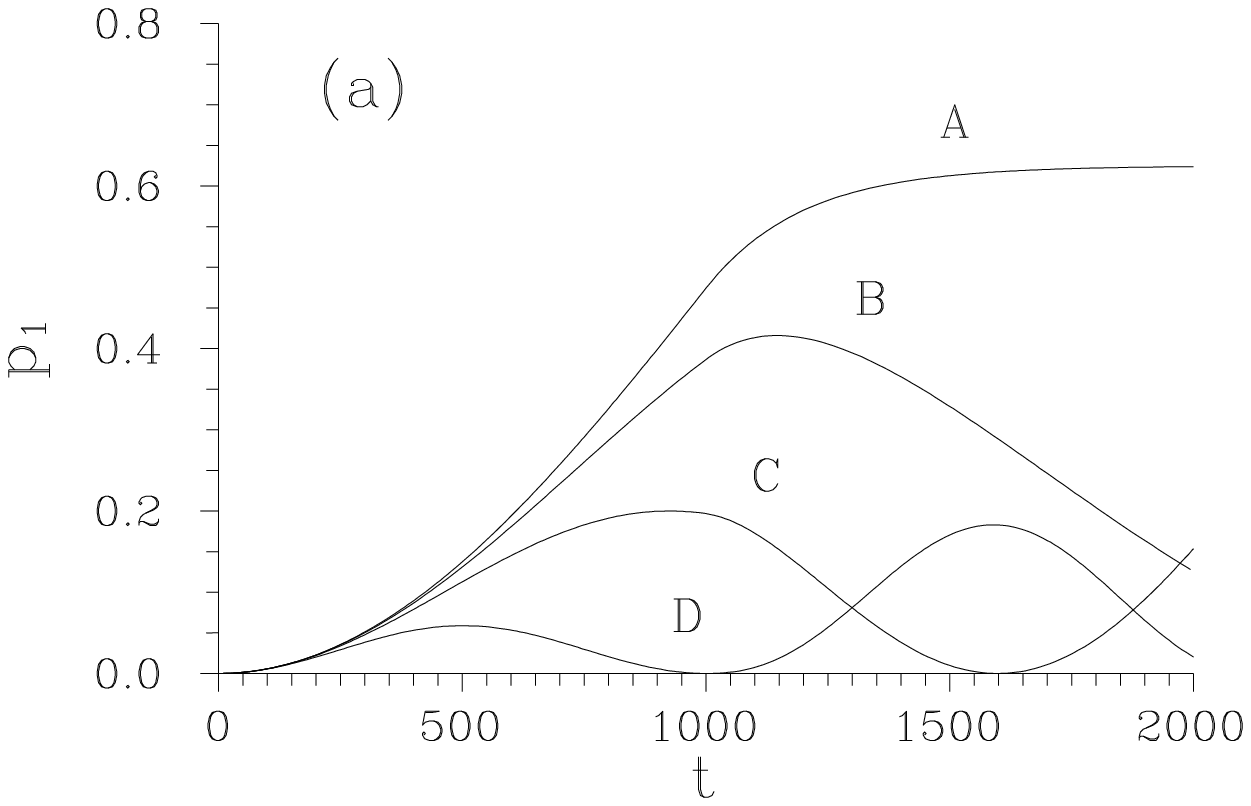,height=4cm}
  \hspace{15mm} \psfig{file=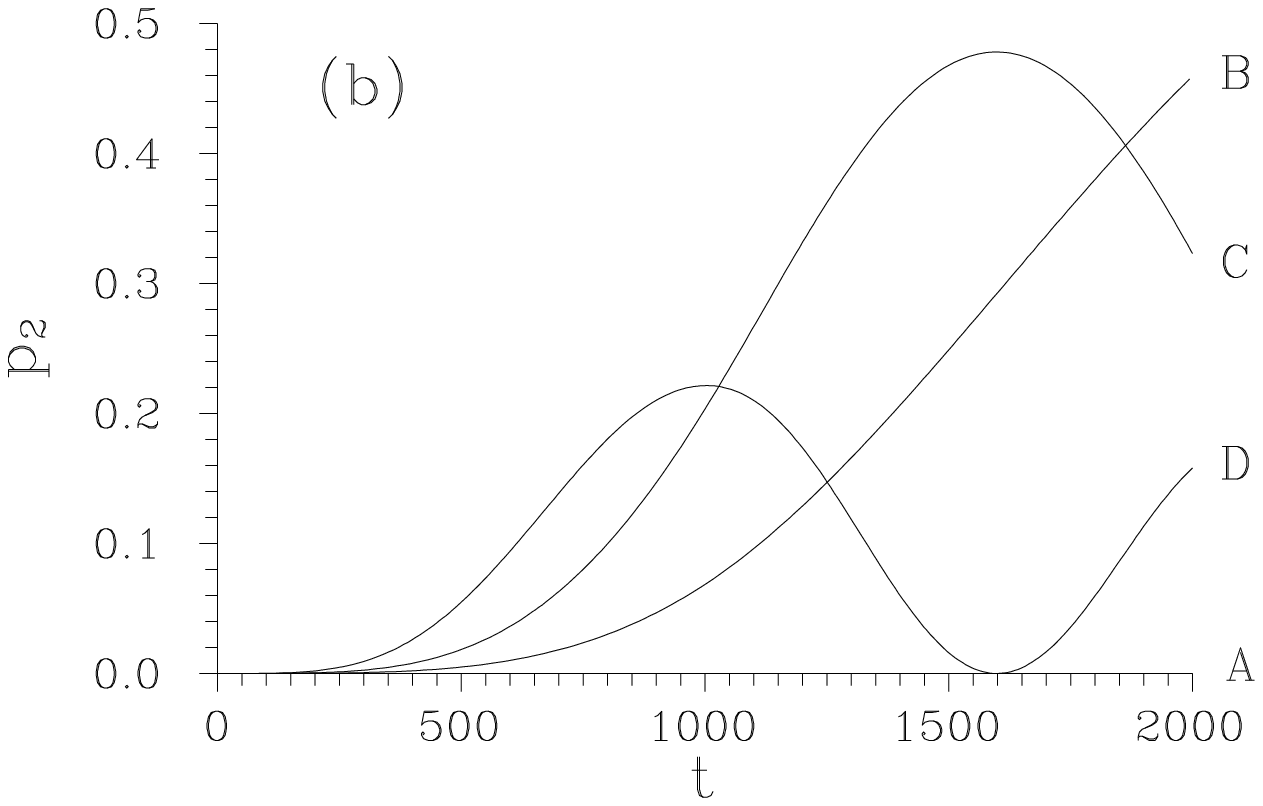,height=4cm} }}
  \vspace{3mm}
  \caption{Increasing values of the transfer integral
  $ J $ influence values of the probabilities $ p_1(t) $ {\bf (a)}
  and $ p_2(t) $ {\bf (b)}; $ J = 10^{-8} $ [A], $ J = 0.0005 $ [B],
  $ J = 0.001 $ [C], $ J = 0.002 $ [D];
  $ F_1 = 0.0005 $, $ F_2 = 0 $, $ \varepsilon = 0 $,
  $ \hbar\delta' = 0 $, $ \tau_1 = 1000 $, $ \tau_2 = 200 $, $ G = 0 $.}
\end{figure}

Molecule 2 can be excited ($ p_2 \ne 0 $) even
in the time when molecule 1 is not excited ($ p_1 = 0 $)
(see Fig.~4 for $ t\approx 300 $). In this time molecule 2
is completely decoupled because $ \rho_r = \rho_i =0 $.
Such an effect has already been observed in
Ref. \cite{tun}. The exciton
subsystem can also return to the excitonless state
($ p_0 = 1 $, see Fig.~4 for $ t \approx 600 $).
However, when the pulse becomes
weak the exciton subsystem cannot reach the state with $ p_0=1 $
(see Figs.~3a,b).
\begin{figure}        % fig. 4
  \centerline{\hbox{\psfig{file=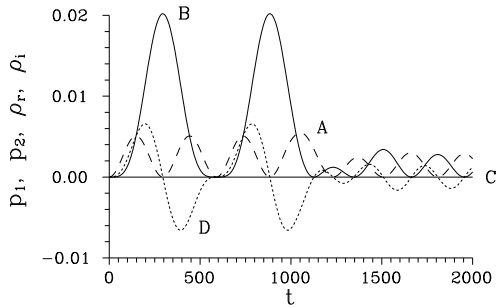,height=4cm} }}
  \vspace{3mm}
  \caption{A typical time development of
  the probabilities $ p_1(t) $ [A] and $ p_2(t) $ [B] and
  the nondiagonal matrix elements $ \rho_r(t) $ [C] and $ \rho_i(t) $ [D]
  in the phononless dimer;
  $ F_1 = 0.0005 $, $ F_2 = 0 $, $ J = 0.007 $, $ \varepsilon = 0 $,
  $ \hbar\delta' = 0 $, $ \tau_1 = 1000 $, $ \tau_2 = 200 $, $ G = 0 $.}
\end{figure}

{\it Interaction with phonons} influences the dynamics as follows.
In case of small $ J $ a decrease of the whole
excitation probability $ p_1(t) + p_2(t) $ (see Figs.~5a,b)
with the increase of $ G $ occurs as a consequence of
the prevailing effect of polaron formation at
molecule 1.
An increase of the whole excitation probability
$ p_1(t) + p_2(t) $ with the increase of $ G $ for greater $ J $
is observed, because $ J $ is renormalized by
the interaction to its smaller values which suppresses
the destructive effect of $ J $ on the excitation creation.
But when $ G $ is great enough (according to the value
of $ J $) the effect of polaron formation at molecule 1 decreases
the whole excitation probability.
The magnitude of $ G $ also influences the time when the
excitation creation is replaced by its annihilation.
Especially, annihilation occurs earlier when the polaron
formation at molecule 1 prevails for smaller $ J $
(compare curves A, B, and C in Figs.~5a,b).
\begin{figure}        % figs. 5a,b
  \centerline{\hbox{ \psfig{file=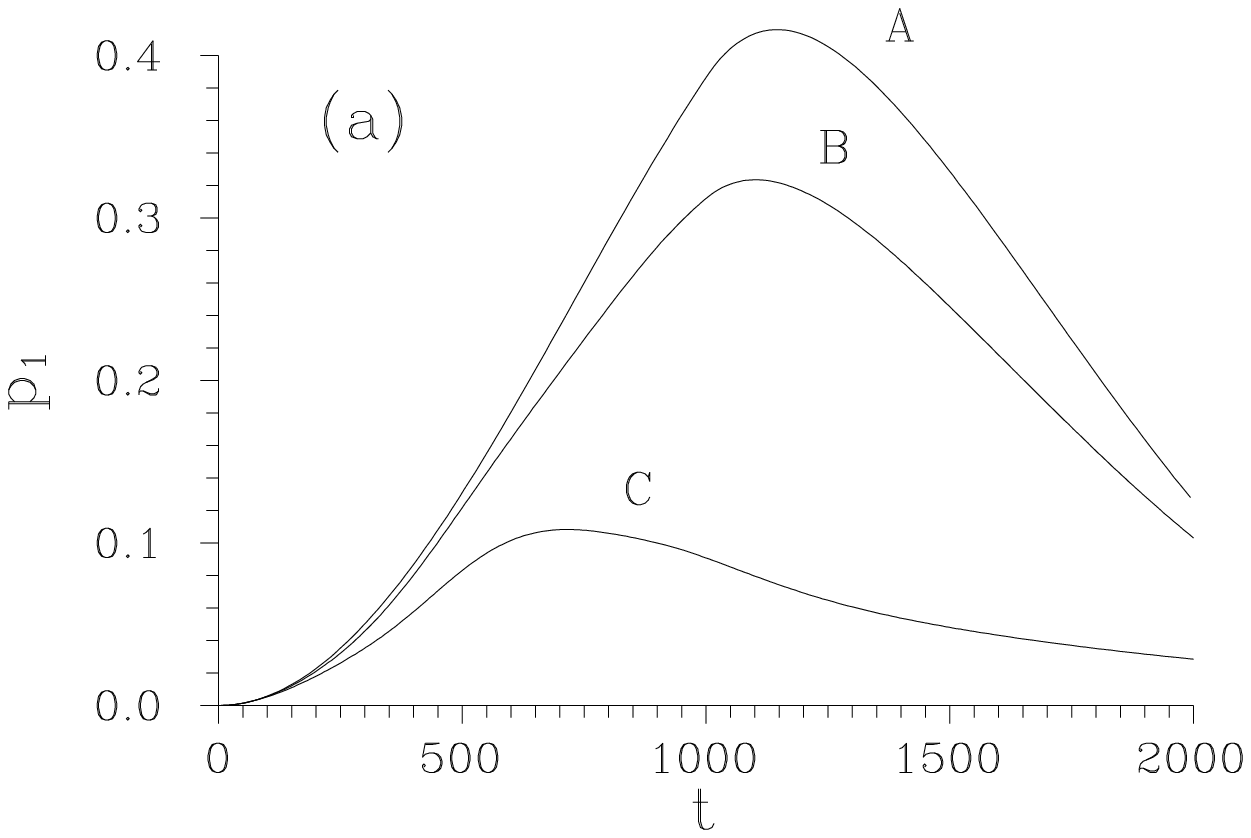,height=4cm}
  \hspace{15mm} \psfig{file=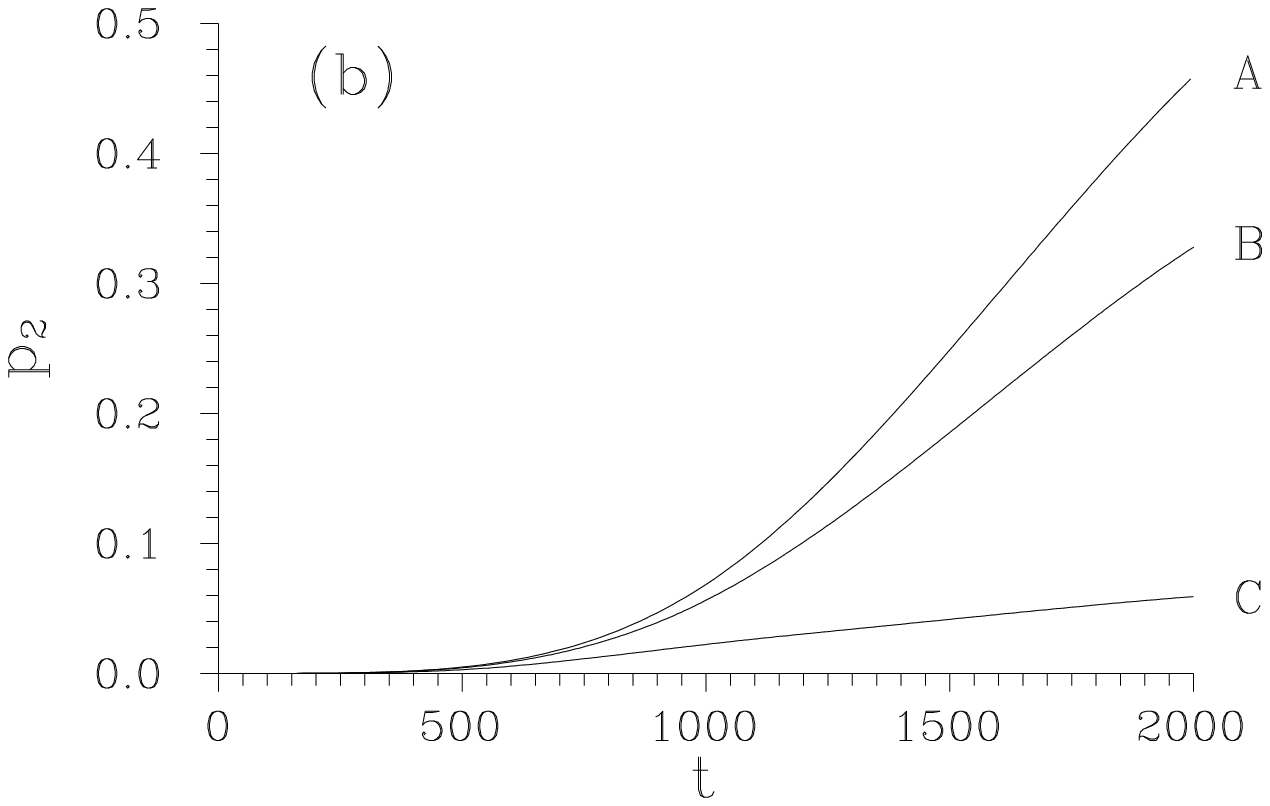,height=4cm} }}
  \vspace{3mm}
  \caption{Increasing values of the exciton--phonon coupling
  constant $ G $ mean the decrease of the probabilities $ p_1(t) $
  {\bf (a)} and $ p_2(t) $ {\bf (b)};
  $ G = 0 $ [A], $ G = 0.003 $ [B], $ G = 0.005 $ [C];
  $ F_1 = 0.0005 $, $ F_2 = 0 $, $ J = 0.0005 $, $ \varepsilon = 0 $,
  $ \hbar\delta' = 0 $, $ \tau_1 = 1000 $, $ \tau_2 = 200 $,
  $ G_1/G = 1+0.25i $, $ G_2/G = 1-0.25i $, $ \hbar\Omega_{ph} = 0.01 $,
  $ \hbar\gamma_{ph} = 0.001 $, $ n_{\rm B} = 0 $.}
\end{figure}

After the pulse is gone, the frequency of excitation exchange
between molecules 1 and 2 decreases
with increasing $ G $. This manifests the renormalization of $ J $.
Also a complete deexcitation of
molecules ($ p_1 = 0 $, $ p_2 = 0 $) cannot be reached
owing to the polaron effect.

\vspace{3mm}
\noindent
{\it 4.2.2. Energetically unbalanced dimer}
\vspace{2mm}

The {\it energy difference} $ 2\varepsilon $ between the excited states
of molecules 1 and 2 affects the phononless system
as follows. Greater
values of the probabilities $ p_1(t) $ (see Fig.~6a) and $ p_1(t) + p_2(t) $
for greater values of $ |\varepsilon| $ have their origin
in decoupling of molecule 2 from
molecule 1 which partially suppresses the effect of coherent
transfer ($ J $). Decoupling of molecule 2 then causes
smaller values of $ p_2(t) $ (see Fig.~6b).
The case when molecule 2 has a lower energy than molecule 1
($ \varepsilon > 0 $, $ \delta' < 0 $) cannot be distinguished
from the case in which molecule 2 has a higher energy than molecule 1
($ \varepsilon < 0 $, $ \delta' > 0 $); time development of $ p_0(t) $,
$ p_1(t) $, and $ p_2(t) $ is the same.
\begin{figure}        % figs. 6a,b
  \centerline{\hbox{ \psfig{file=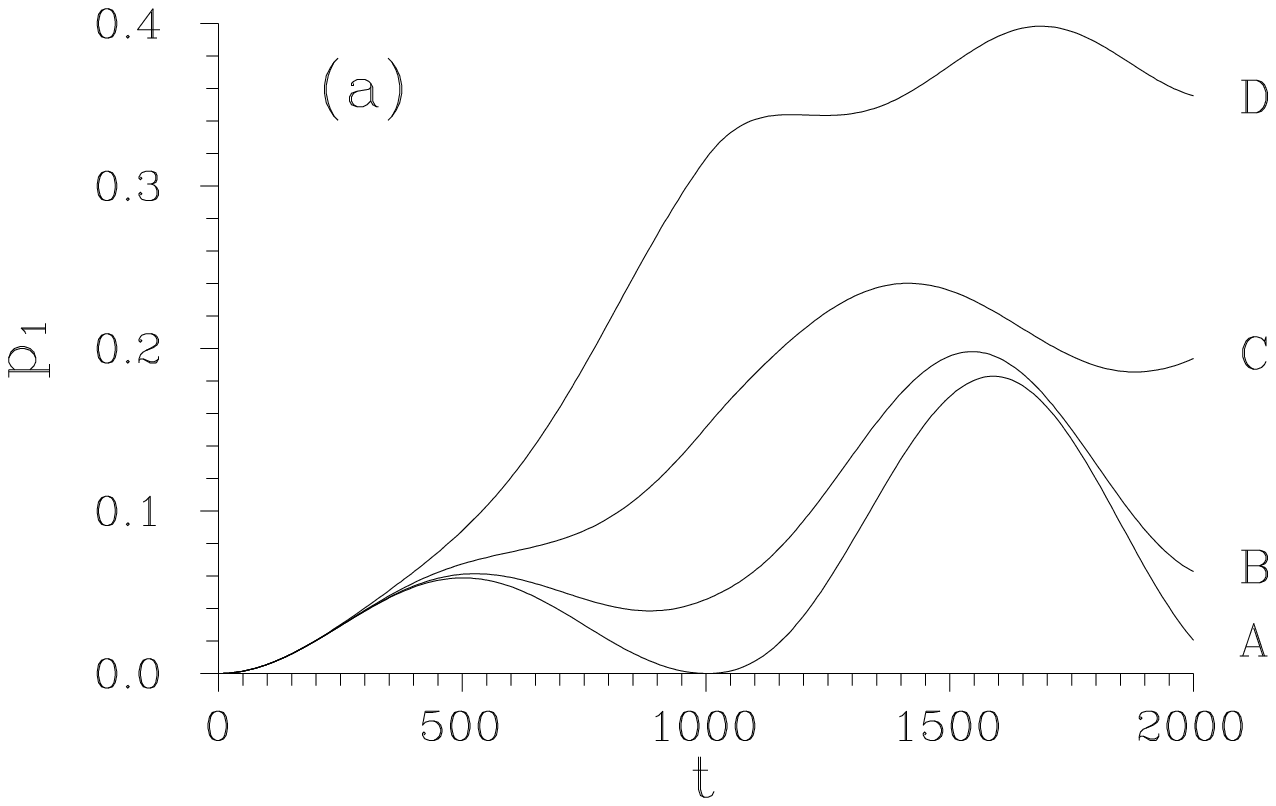,height=4cm}
  \hspace{15mm} \psfig{file=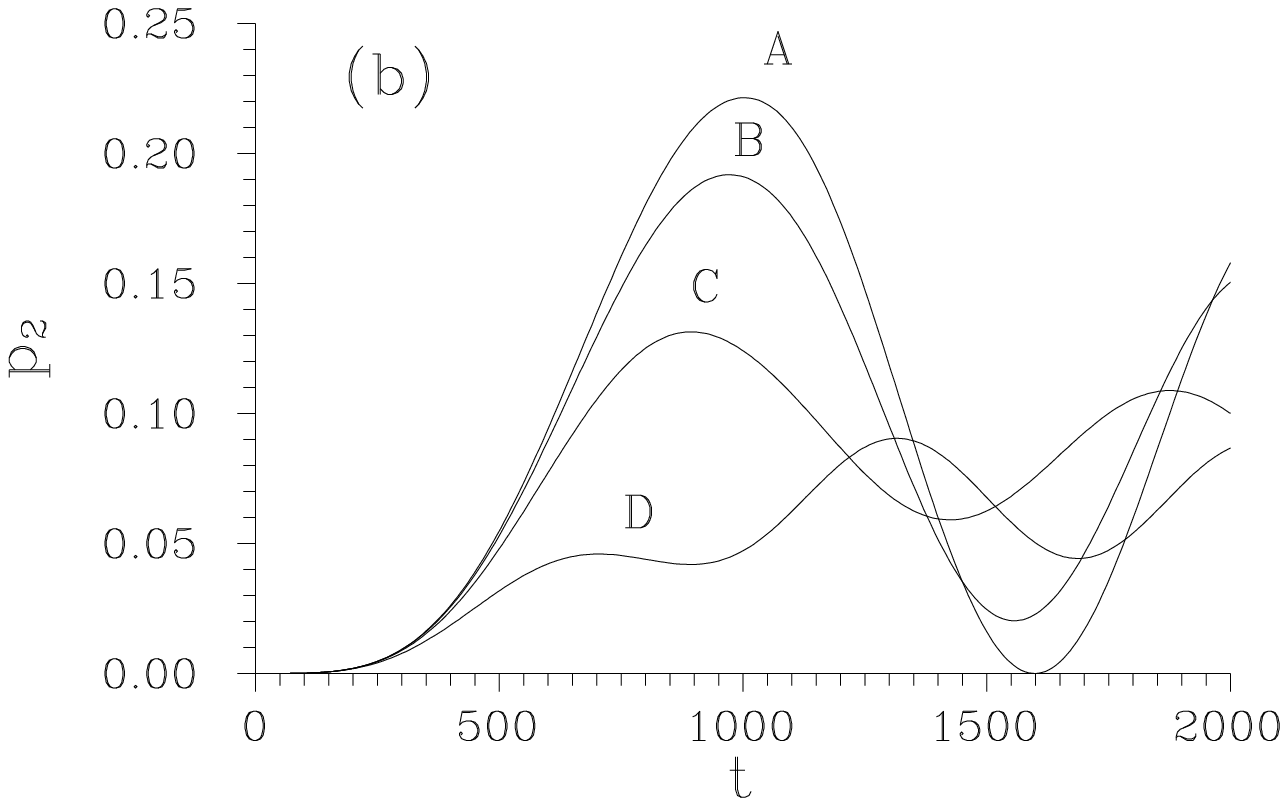,height=4cm} }}
  \vspace{3mm}
  \caption{The increase of the energy difference $ 2\varepsilon $
  suppresses the effect of $ J $, this is demonstrated
  in the time development
  of the probabilities $ p_1(t) $ {\bf (a)} and $ p_2(t) $
  {\bf (b)};
  $ \varepsilon = 0 $, $ \hbar\delta' = 0 $ [A], $ \varepsilon = 0.0005 $,
  $ \hbar\delta' = -0.0005 $ [B],
  $ \varepsilon = 0.001 $, $ \hbar\delta' = -0.001 $ [C],
  $ \varepsilon = 0.002 $, $ \hbar\delta' = -0.002 $ [D];
  $ F_1 = 0.0005 $, $ F_2 = 0 $, $ J = 0.002 $,
  $ \tau_1 = 1000 $, $ \tau_2 = 200 $,
  $ G = 0 $.}
\end{figure}

The increase of $ \varepsilon $ may result in the decrease
of the probability $ p_1(t) + p_2(t) $
when {\it exciton--phonon coupling} is nonzero.
Also the above mentioned indistinguishability is lost.

If {\it molecule 2 has a lower energy
than molecule 1} (positive $ \varepsilon $), values of the
probabilities $ p_1(t) $ (see Fig.~7a) and $ p_1(t) + p_2(t) $
(compare Figs.~7a,b) increase with the increase of $ G $.
This indicates the increasing renormalization of $ J $.
However, when values of $ G $ are too great, then polaron formation at
molecule 1 enters
into play and the lowering of values of $ p_1(t) $ and $ p_1(t) + p_2(t) $
is observed.

Increasing positive $ \varepsilon $ leads to greater values
of $ p_1(t) + p_2(t) $ in cases when the polaron formation at
molecule 1 does not play an important role
(for smaller values of $ G $). This is caused by the decrease of
the destructive effect of transfer on the excitation creation.
However, the increase of positive $ \varepsilon $ means
the decrease of $ p_1(t) + p_2(t) $ if the polaron formation
at molecule 1 plays a dominant role. In this case,
the destructive effect of transfer on the polaron formation
at molecule 1 becomes weaker with increasing $ \varepsilon $.

\begin{figure}        % figs. 7a,b
  \centerline{\hbox{ \psfig{file=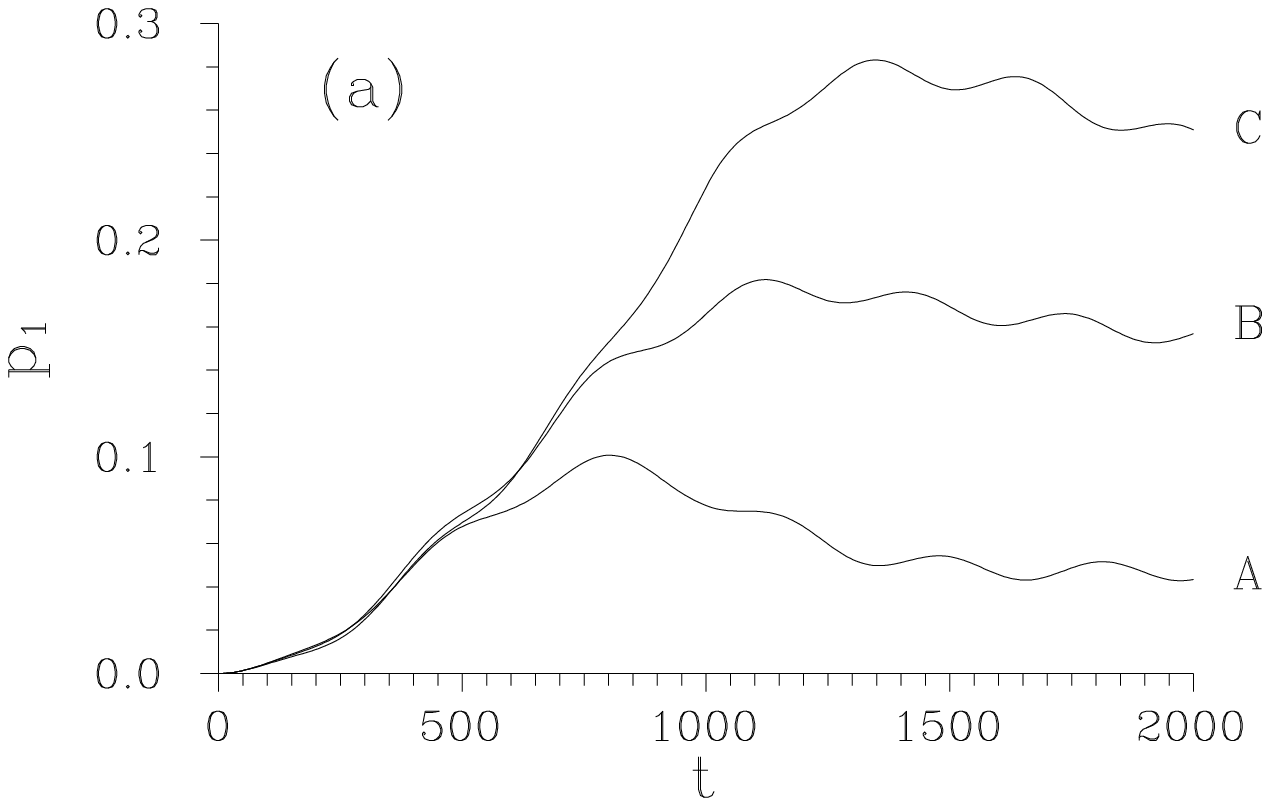,height=4cm}
  \hspace{15mm} \psfig{file=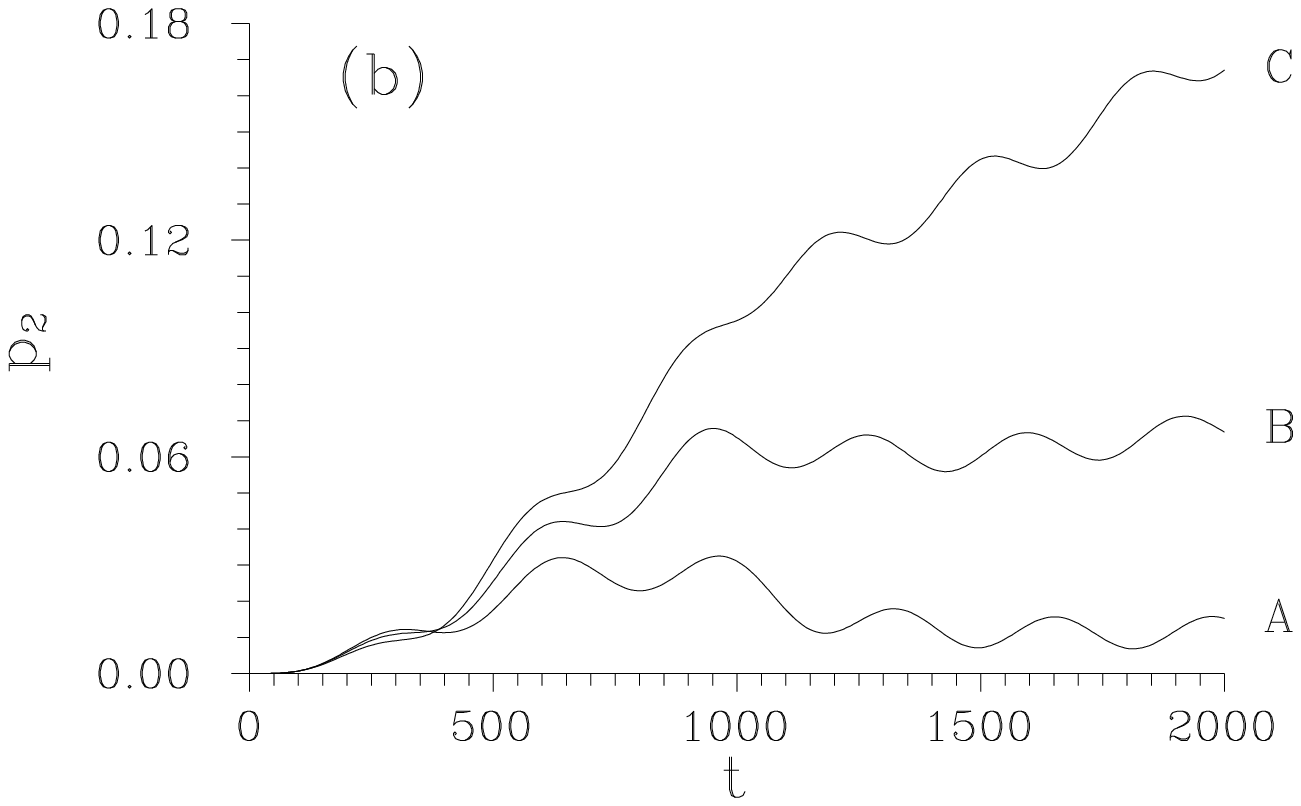,height=4cm} }}
  \vspace{3mm}
  \caption{Increasing values of the exciton--phonon coupling constant
  $ G $ suppress the effect of $ J $ and that means the increase
  of the probabilities $ p_1(t) $ {\bf (a)}
  and $ p_2(t) $ {\bf (b)};
  $ G = 0 $ [A], $ G = 0.003 $ [B], $ G = 0.005 $ [C];
  $ F_1 = 0.0005 $, $ F_2 = 0 $, $ J = 0.005 $, $ \varepsilon = 0.004 $,
  $ \hbar\delta' = -0.004 $, $ \tau_1 = 1000 $, $ \tau_2 = 200 $,
  $ G_1/G = 1+0.25i $, $ G_2/G = 1-0.25i $, $ \hbar\Omega_{ph} = 0.01 $,
  $ \hbar\gamma_{ph} = 0.001 $, $ n_{\rm B} = 0 $.}
\end{figure}

If {\it molecule 2 has a higher energy than molecule 1}
(negative $ \varepsilon $),
the increase of $ G $ results in the decrease of the probabilities $ p_1(t) $
and $ p_1(t) + p_2(t) $ in all cases. Thus the interaction with
phonons supports the destructive effect of $ J $ on the
excitation creation.

The increase of $ J $ need not mean only smaller values of $ p_1(t) $
caused by the destructive effect of $ J $ on the excitation
creation. Especially, when a strong interaction with
phonons forms a polaron at molecule 1
suppressing its excitation, greater values of $ J $ can partially break the
polaron formation and admit greater values of $ p_1(t) $
(this effect is well pronounced for greater
positive $ \varepsilon $).

An exciton moving on the dimer can be ``partially localized''
\cite{Agra1} by the interaction with phonons
(see Fig.~8). A complete localization of the exciton cannot be reached
because coherent transfer cannot be completely
suppressed (see asymptotic expressions in Section~6).
\begin{figure}        % fig. 8
  \centerline{\hbox{\psfig{file=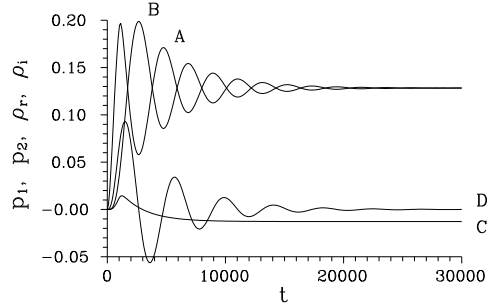,height=4cm} }}
  \vspace{3mm}
  \caption{The ``partial localization'' of exciton in the dimer,
  described by the probabilities $ p_1(t) $ [A] and $ p_2(t) $ [B] and
  by the nondiagonal matrix elements $ \rho_r(t) $ [C] and $ \rho_i(t) $ [D];
  $ F_1 = 0.0005 $, $ F_2 = 0 $, $ J = 0.0005 $, $ \varepsilon = 0 $,
  $ \hbar\delta' = 0 $, $ \tau_1 = 1000 $, $ \tau_2 = 200 $,
  $ G = 0.005 $,
  $ G_1/G = 1+0.25i $, $ G_2/G = 1-0.25i $, $ \hbar\Omega_{ph} = 0.01 $,
  $ \hbar\gamma_{ph} = 0.01 $, $ n_{\rm B} = 0 $.}
\end{figure}

\vspace{3mm}
\noindent
{\it 4.3. Dimer with both molecules interacting with a pulse}
\vspace{2mm}

We first address the validity of
``single-excitation'' approximation
considering a two-level system
and a symmetric dimer with $ J $ being practically zero
and comparing their levels of excitation.
Omission of the two-exciton state in the dimer
manifests itself in lower values of the excitation probability
$ p_1(t) $ (or $ p_2(t) $) in comparison with that for
a two-level system. The deviations
can be estimated for a given value of the excitation probability
from curves in Fig.~9.
In general, the ``single-excitation''
approximation is very well accepted
for values of the whole excitation probability
$ p_1(t) + p_2(t) $ up to $ \approx 0.2 $. The deviations in
$ p_1(t) + p_2(t) $ are less than
about 10~\% for values of $ p_1(t) + p_2(t) $ up to $ \approx 0.6
$, which is also well acceptable.

Results of this section are not applicable to the electron
dynamics because only one excitationless state
for the whole system has been considered.

\begin{figure}        % fig. 9
  \centerline{\hbox{\psfig{file=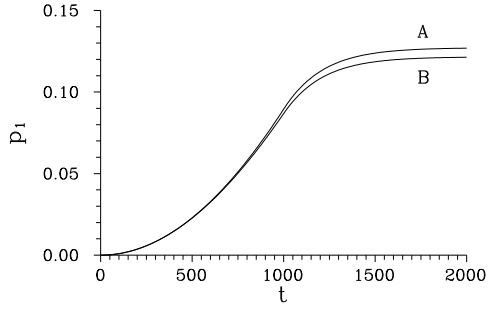,height=4cm} }}
  \vspace{2mm}
  \caption{The time development of the probability $ p_1(t) $
  shows the accuracy of the ``single-excitation''
  approximation; $ F_2 = 0 $ [A], $ F_2 = 0.0002 $ [B];
  $ F_1 = 0.0002 $, $ J = 10^{-8} $, $ \varepsilon = 0 $,
  $ \hbar\delta' = 0 $, $ \tau_1 = 1000 $, $ \tau_2 = 200 $,
  $ G = 0 $.}
\end{figure}

\vspace{3mm}
\noindent
{\it 4.3.1. Energetically balanced dimer}
\vspace{2mm}

{\it Coherent transfer} ($ J $) affects the excitation dynamics
similarly as it is discussed in
Subsection 4.2; i.e. the increase of $ J $ in the phononless system
means the decrease of the whole excitation probability $ p_1(t) +
p_2(t) $. The exciton subsystem can return to the excitonless
state ($ p_0 = 1 $).

The increasing {\it interaction with phonons} ($ G $) decreases
$ p_1(t) + p_2(t) $ for small $ J $ as
a result of the polaron formation at both molecules.
However, the increase of $ G $ leads to the increase of
$ p_1(t) + p_2(t) $ for greater values
of $ J $ because the destructive effect of transfer on the excitation
creation is partially suppressed.

\vspace{3mm}
\noindent
{\it 4.3.2. Energetically unbalanced dimer}
\vspace{2mm}

We assume that molecule 1 is pumped
resonantly and molecule 2 nonresonantly in the further discussion.
Despite the effect of coherent transfer, small values of
$ \varepsilon $ lead to a strong
asymmetry in the time development of the probabilities $ p_1(t) $
and $ p_2(t) $ (see Fig.~10) having its origin in
the nonresonant interaction of molecule 2 with the pulse.
\begin{figure}        % fig. 10
  \centerline{\hbox{\psfig{file=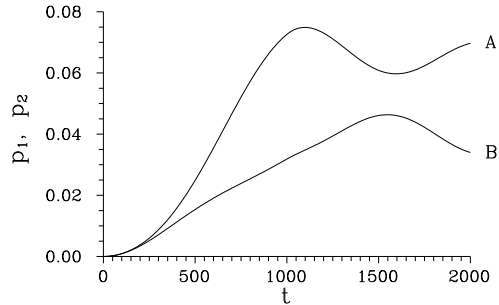,height=4cm} }}
  \vspace{2mm}
  \caption{A strong asymmetry in the time development of
  the probabilities $ p_1(t) $ [A] and
  $ p_2(t) $ [B] is caused by nonzero $ \varepsilon $;
  $ F_1 = 0.0002 $, $ F_2 = 0.0002 $, $ J = 0.002 $, $ \varepsilon = 0.0005 $,
  $ \hbar\delta' = -0.0005 $, $ \tau_1 = 1000 $, $ \tau_2 = 200 $,
  $ G = 0 $.}
\end{figure}
The energy position of molecule 2 with respect to molecule 1
(the sign of $ \varepsilon $) strongly influences
the excitation dynamics (see Fig.~11);
the probability $ p_1(t) $ reaches
greater values for molecule 2 having a lower energy than molecule 1.
Increasing $ |\varepsilon| $ smooths this asymmetry.
\begin{figure}        % fig. 11
  \centerline{\hbox{\psfig{file=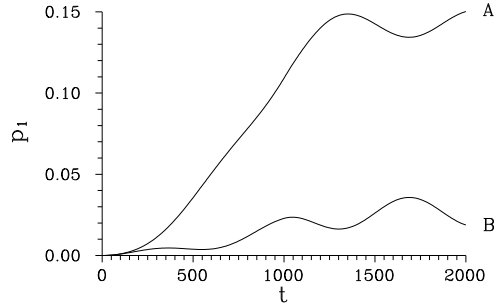,height=4cm} }}
  \vspace{2mm}
  \caption{The sign of the energy difference $ 2\varepsilon $
  strongly influences the time development of the
  probability $ p_1(t) $; $ \varepsilon = 0.002 $,
  $ \hbar\delta' = -0.002 $ [A],
  $ \varepsilon = -0.002 $, $ \hbar\delta' = 0.002 $ [B];
  $ F_1 = 0.0002 $, $ F_2 = 0.0002 $, $ J = 0.002 $,
  $ \tau_1 = 1000 $, $ \tau_2 = 200 $,
  $ G = 0 $.}
\end{figure}

{\it Interaction with phonons} modifies the excitation dynamics
as follows. For {\it molecule 2 having a lower energy than molecule
1} ($ \varepsilon > 0 $), the probability $ p_1(t) + p_2(t) $
increases with increasing $ G $ until the range
of values of $ G $ is reached in which the polaron formation at
molecule 1 causes the decrease of $ p_1(t) + p_2(t) $. The higher
the $ \varepsilon $ is the less the destructive effect of $ J $
on the polaron
formation at molecule 1 is.
The excitation dynamics is then influenced for weaker $ G $.

When the {\it energy of molecule 2 is} slightly {\it higher than that
of molecule 1} ($ \varepsilon < 0 $),
the increase of $ G $ leads to the
decrease of $ p_1(t) $ (effects of transfer and polaron formation
support each other). But an increase of $ p_2(t) $ occurs
because interaction with phonons tunes molecule 2 to the resonance
with the field. The probability $ p_1(t) + p_2(t) $ also increases
with increasing $ G $.
For greater values of $ |\varepsilon| $, the probability
$ p_1(t) + p_2(t) $ decreases with increasing $ G $.

After the pulse is gone the excitation dynamics is the same
as that discussed in Subsection~4.2.

\vspace{6mm}

The obtained numerical results elucidate the validity of
perturbation approximation.
We consider the perturbation approximation in the exciton--photon
coupling constants to be reliable for arbitrary times at least for classical
deterministic fields.
The positive semidefiniteness of
the exciton reduced density matrix has been slightly
breaked for longer times and greater values of $ G $. This
clearly indicates that it is caused by the omission
of terms proportional to the
third and higher powers of the exciton--phonon coupling
constants.
However, deviations are small for real values of
parameters and the perturbation approximation in
the exciton--phonon coupling constants is also justified.

\vspace{4mm}
\noindent
{\Large \bf 5. Influence of optical field fluctuations}
\vspace{3mm}

Effects originating both in the amplitude and phase fluctuations
of a pulsed optical field on the excitation dynamics are studied.

We assume that the noisy part of the photon field amplitude is
proportional to the strength of its coherent part.
The normalized photon field correlation function
$ \delta N_n(t,\tau) $ has then the form (the strength
of the field is absorbed into $ F_1 $ and $ F_2 $):
\begin{equation}             % 26
  \delta N_n(t,\tau) = \tilde{\cal A}_n(t) \tilde{\cal A}^*_n(\tau) \langle
  \delta {\cal A}_n(t) \delta {\cal A}^*_n(\tau)\rangle_f ,
\end{equation}
where the normalized envelope of the pulse is given in Eq. (25).
The optical field fluctuations are described by the amplitude
deviation $ \delta{\cal A}_n(t) $,
\begin{equation}             % 27
 \delta{\cal A}_n(t) = \frac{1}{\tilde{\cal A}(t)} \sum_{K} \sqrt{
 \frac{\hbar}{ 2\epsilon_0V\omega_K} } \delta\alpha_K(t) ,
\end{equation}
normalized with respect to the envelope $ \tilde{\cal A}(t) $
defined in (A9) in Appendix.
A stochastic model (\cite{Per}, p.~137) for both the amplitude and
phase fluctuations of $ \delta{\cal A}_n(t) $
provides the relation
\begin{equation}             % 28
 \langle \delta {\cal A}_n(t) \delta{\cal A}^*_n(\tau)\rangle_f =
  \langle|\delta
 {\cal A}_n|^2\rangle_f \exp\left[ -i(\omega_{K_0} + \omega_s)(t-\tau) \right]
 \langle \exp\left[ i\phi(t) -i\phi(\tau) \right] \rangle_f .
\end{equation}
The moment $ \langle|\delta{\cal A}_n|^2\rangle_f $ involves
averaging over amplitude fluctuations,
$ \omega_s $ denotes the frequency shift of amplitude fluctuations
and the factor $ \langle\exp\left[ i\phi(t) -
i\phi(\tau)\right] \rangle_f $
involves averaging over the phase $ \phi(t) $. If $ \phi(t) - \phi
(\tau) = \int^t_{\tau} d\tau' \, \Delta\omega(\tau') $, where $ \Delta
\omega(\tau) $ represents the Gaussian Markovian process, then
Eq. (28) can be rewritten into the form
\begin{equation}             % 29
 \langle \delta {\cal A}_n(t) \delta{\cal A}^*_n(\tau)\rangle_f = n_s
 \exp\left[ -i(\omega_{K_0} + \omega_s)(t-\tau) \right] \exp\left[ -\gamma_s
 |t-\tau| \right] .
\end{equation}
The quantity $ n_s = \langle|\delta{\cal A}_n|^2\rangle_f $
characterizes the
strength of noise, whereas $ \gamma_s $ describes the strength of
phase correlations ($ \langle\Delta\omega(t)\Delta\omega(t')\rangle =
2\gamma_s \delta(t-t') $).

The correlation
function $ \delta N_n(t,\tau) $ in (26) then gets the form:
\begin{equation}             % 30
 \delta N_n(t,\tau) = \tilde{\cal A}_n(t) \tilde{\cal A}^*_n(\tau) n_s
 \exp\left[ -i(\omega_{K_0} + \omega_s)(t-\tau) \right] \exp\left[ -\gamma_s
 |t-\tau| \right] .
\end{equation}

We further omit vacuum fluctuations in the photon field and thus
$ \delta N_v(t,\tau) = \delta N(t,\tau) $ (see Eq. (14)).
We also assume $ \delta N_a(t,\tau) = 0 $
as a consequence of averaging over the phases of
$ \delta\alpha^*_K(t) $ and $ \delta\alpha^*_{K'}(\tau) $
in $ \langle\delta\alpha^*_K(t)\delta
\alpha^*_{K'} (\tau) \rangle_f $ in Eq. (14).

The influence of noise on the excitation dynamics is described by
the matrices $ {\cal F}_1(t) $ and $ {\cal F}_4(t) $ given in
Eq. (A7) in Appendix.
The time-dependent coefficients $ M_1(t) = \bar{M}_1(t), \ldots, P_2(t) =
\bar{P}_2(t) $ defined in Eqs. (A10---A12) contain the normalized
correlation function $ \delta N_n(t,\tau) $ given in Eq. (30) instead of
$ \delta N(t,\tau) $.

Optical noise influences the dynamics of a two-level system
in such a way that
it forces the system to evolve in the direction to the state with
$ p_0 = p_1 = 1/2 $, $ \rho_{10} = 0 $ (see Fig.~12).
The increase of $ n_s $ leads to the increased influence
of noise. The decrease of $ \gamma_s $ means the increase of the
correlation time
of noise $ \tau_s $ ($ \tau_s = 1/\gamma_s $) and results in the
increased effect of noise.
\begin{figure}        % fig. 12
  \centerline{\hbox{\psfig{file=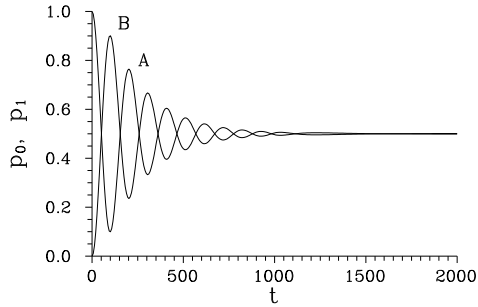,height=4cm} }}
  \vspace{3mm}
  \caption{A typical excitation dynamics in the two-level system
  interacting with a noisy pulse described by the probabilities
  $ p_0(t) $ [A] and $ p_1(t) $ [B];
  $ F_1 = 0.01 $, $ F_2 = 0 $, $ J = 10^{-8} $, $ \varepsilon = 0 $,
  $ \hbar\delta' = 0 $, $ \tau_1 = 1000 $, $ \tau_2 = 200 $, $ G = 0 $,
  $ n_s = 0.1 $, $ \gamma_s = 0.01 $, $ \omega_s = 0 $.}
\end{figure}

For values of the exciton-photon coupling constants
$ F_1 $ and $ F_2 $ and pulse durations ($ \tau_1 + \tau_2 $)
used in Section 4, the influence of noise with
reasonable values of parameters ($ n_s \leq 0.1 $,
$ \gamma_s \geq 0.01 $) is negligible.

These results show that
noise in ultrashort pulsed optical fields does not influence
substantially experimental results under standard conditions.
Effects originating in nonclassical properties of photon fields
are the matter of further investigations.

\vspace{4mm}
\noindent
{\Large \bf 6. Long-time behaviour}
\vspace{3mm}

Eq. (20) is analyzed for times when
the photon field (the pulse) does not act
on the exciton subsystem
and for times longer than the relaxation
time $ \tau_R $ of the phonon reservoir.
Vacuum fluctuations of the photon field
are also omitted and hence, e.g., the exciton
decay processes are not taken into account.
Then the third matrix in Eq. (20) is zero
and the sets of equations for $ {\cal R}_1(t) $ and $ {\cal R}_2(t) $
are independent.
The time-dependent coefficients in the matrices $ {\cal G}_1(t) $ and
$ {\cal G}_2(t) $ can be
replaced by their asymptotic values for $ t-t_0\rightarrow\infty $.

The excitation dynamics is then
driven by the following equations:
\begin{equation}             % 31
 \frac{d}{dt} \left[
  \begin{array}{c} \rho_{11}(t) \\ \rho_{22}(t) \\ \rho_{r}(t) \\
   \rho_{i}(t) \end{array} \right]
 = \frac{1}{\hbar} \left[
  \begin{array}{cccc}
     0 & 0 & 0 & -2J \\
     0 & 0 & 0 & 2J \\
     0 & 0 & 0 & 2\varepsilon \\
     J & -J & -2\varepsilon & 0
  \end{array} \right] \left[
  \begin{array}{c} \rho_{11}(t) \\ \rho_{22}(t) \\ \rho_{r}(t) \\
   \rho_{i}(t) \end{array} \right] - \left[
  \begin{array}{cccc}
     0 & 0 & 0 & 0 \\
     0 & 0 & 0 & 0 \\
     A_{\rm as} & C_{\rm as} & E_{\rm as} & -F_{\rm as} \\
     B_{\rm as} & -D_{\rm as} & F_{\rm as} & E_{\rm as}
  \end{array} \right] \left[
 \begin{array}{c} \rho_{11}(t) \\ \rho_{22}(t) \\ \rho_{r}(t) \\
   \rho_{i}(t) \end{array} \right] .
\end{equation}
The coefficients $ A_{\rm as}, \ldots, F_{\rm as} $
express asymptotic values of
the time-dependent coefficients $ A(t),$ $ \ldots, F(t) $ given
in (A3) in Appendix for
$ t-t_0 \rightarrow\infty $;
\begin{eqnarray}             % 32
 A_{\rm as} &=& \frac{\pi J}{\hbar N} \sum_k |G^1_k - G^2_k|^2 \left\{
  \varepsilon \left[ 2n_{\rm B}(\hbar\Omega_k) +1 \right] + \Delta \right\}
  \delta(\hbar \Omega_k-2\Delta) , \nonumber \\
 B_{\rm as} &=& \frac{J}{\hbar N} \sum_k |G^1_k - G^2_k|^2 {\cal P}' \frac{
  \Omega^2_k}{\Omega^2_k -4\Delta'^2} \left\{ \left[ 2n_{\rm B}(\hbar\Omega_k)
  +1 \right] + \frac{2\varepsilon}{\hbar\Omega_k} \right\} , \nonumber \\
 C_{\rm as} &=& \frac{\pi J}{\hbar N} \sum_k |G^1_k - G^2_k|^2 \left\{
  -\varepsilon \left[ 2n_{\rm B}(\hbar\Omega_k) +1 \right] + \Delta
  \right\} \delta(\hbar \Omega_k-2\Delta) , \nonumber \\
 D_{\rm as} &=& \frac{J}{\hbar N} \sum_k |G^1_k - G^2_k|^2 {\cal P}' \frac{
  \Omega^2_k}{\Omega^2_k -4\Delta'^2} \left\{ \left[ 2n_{\rm B}(\hbar\Omega_k)
  +1 \right] - \frac{2\varepsilon}{\hbar\Omega_k} \right\} , \nonumber \\
 E_{\rm as} &=& \frac{2\pi J^2}{\hbar N} \sum_k |G^1_k - G^2_k|^2
  \left[ 2n_{\rm B}(\hbar\Omega_k) +1 \right] \delta(\hbar \Omega_k-2\Delta) ,
  \nonumber \\
 F_{\rm as} &=& - \frac{1}{N} \sum_k \left( G^1_k - G^2_k \right)
  \left( G^1_{-k} + G^2_{-k} \right) \Omega_k ,
\end{eqnarray}
$ {\cal P}' $ denotes principal value.

For $ \varepsilon = 0 $ (it implies that $ B_{\rm as} = D_{\rm as} $) and
$ J \ll
\Delta W_{ph} $ ($ \Delta W_{ph} $ being the width of phonon spectrum)
the transfer integral $ J $ at the positions (4,1) and (4,2)
in Eq. (31) is renormalized
to a new value
\begin{eqnarray}             % 33
 J_{ren} = J -\hbar B_{\rm as} = J\exp(-2W) + o(G^2) , \nonumber \\
 W = \frac{1}{2N} \sum_k |G^1_k - G^2_k|^2 \left[ 2n_{\rm B}(\hbar\Omega_k)
  +1 \right] ,
\end{eqnarray}
where $ W $ denotes the Debye-Waller factor describing the small polaron
formation \cite{Hol,Cap2}.

The transfer integral $ J $ at the positions (4,1) and
(4,2) is renormalized twice and $ J $ at the positions (1,4) and (2,4) is
not renormalized. But the effect of renormalization on the
matrix elements $ \rho_{11}(t) $ and $ \rho_{22}(t) $ is correct
as can be shown from the equations for $ \rho_{11}(t) $ and
$ \rho_{22}(t) $ with the nondiagonal matrix elements $ \rho_{12}(t) $
and $ \rho_{21}(t) $ excluded \cite{Cap2}.

The asymptotic state of the dimer (i.e. the state reached
for $ t - t_0 \rightarrow \infty $) has the form
correct up to $ G^2 $ as follows,
\begin{equation}             % 34
 \left( \rho^{\rm as}_{11}, \rho^{\rm as}_{22},
 \rho^{\rm as}_{r},
 \rho^{\rm as}_{i} \right) \approx \left( -\gamma_1\Delta +
 \gamma_2\varepsilon,
 -\gamma_1\Delta - \gamma_2\varepsilon, J\gamma_2, 0 \right) ,
\end{equation}
where
\begin{eqnarray}             % 35
 \gamma_1 &=& \frac{1}{N} \sum_k |G^1_k - G^2_k|^2 \left[
  2n_{\rm B}(\hbar\Omega_k) +1 \right] \delta(\hbar\Omega_k-2\Delta) ,
  \nonumber \\
 \gamma_2 &=& \frac{1}{N} \sum_k |G^1_k - G^2_k|^2 \delta
  (\hbar\Omega_k-2\Delta) .
\end{eqnarray}
Its form written in the basis of
the eigenstates $ |+\rangle $ and $ |-\rangle $ (17),
\begin{equation}             % 36
 \frac{\rho^{\rm as}_{++} }{ \rho^{\rm as}_{--}} = \frac{\Delta -
  \varepsilon}{\Delta + \varepsilon} \exp(-2\beta\Delta) ,
 \hspace{2cm}
 \rho^{\rm as}_{+-} = \rho^{\rm as}_{-+} = 0 ,
\end{equation}
is in agreement with the result of equilibrium statistical
physics for an already created excitation.
This indicates that the theory
is valid also for longer times despite the perturbation
approximation applied in the derivation.

\vspace{4mm}
\noindent
{\Large \bf 7. Conclusions}
\vspace{3mm}

A microscopic theory of the excitation dynamics in molecular
condensates interacting with optical
fields (with ultrashort pulses) has been developed.
It has provided a tool for the investigation of the initial
stage of excitation dynamics, i.e. when an excitation
is being created.

The theory has been applied to a dimer under the conditions
when times characterizing pulse duration and propagation and
relaxation of an excitation are comparable.
The following results have been obtained.
In this case, coherent transfer strongly influences the
processes of excitation
creation and annihilation. The stronger the coherent transfer
is, the lower the excitation level is.
The influence of coherent transfer is suppressed
by the increase of energy difference between
excited states.
Interaction with phonons leads in general
to polaron formation. This leads to lower excitation levels
on one side, but on the other side it
partially preserves the excitation from being annihilated.
Interesting behaviour occurs when effects of
coherent transfer and polaron formation compete.
Interaction with phonons can partially suppress the destructive
effect of coherent transfer on the excitation creation; this
leads to higher excitation levels. This occurs
when the energy of a molecule being excited is higher
than that of surrounding molecules.
On the other hand, interaction with phonons can support
the effect of coherent transfer on the excitation creation
and thus can lead to the further lowering of the excitation level.
This is the case when the energy of a molecule being excited
is lower than that of surrounding molecules.
The theory also
provides a long-time dynamics being in agreement with
equilibrium statistical physics.

It has been shown that noise in an ultrashort pulse
does not influence substantially the initial stage of excitation
dynamics under standard conditions.
However, the theory is also suitable for the description
of effects originating in nonclassical properties of
interacting
optical fields (e.g., those with squeezed vacuum fluctuations).

\vspace{4mm}
\noindent
{\Large \bf Acknowledgements}
\vspace{3mm}

The author thanks V. \v{C}\'{a}pek for the suggestion of the
theme, his advice, and discussions.
He also thanks J.~Pe\v{r}ina for advice
and discussions.

\newpage
\appendix
\section{}
\noindent
{\Large \bf Definitions of matrices in Eq. (20)}
\vspace{3mm}

Appendix contains the definitions of the
matrices $ {\cal J}_1 $, $ {\cal J}_2 $, $ {\cal G}_1(t) $,
$ {\cal G}_2(t) $, $ {\cal F}_1(t) $,
$ {\cal F}_2(t) $, $ {\cal F}_3(t) $, and $ {\cal F}_4(t) $
in Eq. (20).
The influence of their elements on the
excitation dynamics is also discussed.

The matrices $ {\cal J}_1 $ and $ {\cal J}_2 $ describe the
dynamics of the free exciton system:
\begin{eqnarray}             % A1
 {\cal J}_1 = \frac{1}{\hbar} \left[ \begin{array}{ccccc}
   0 & 0 & 0 & -2J & 0 \\
   0 & 0 & 0 & 2J & 0 \\
   0 & 0 & 0 & 2\varepsilon & 0 \\
   J & -J & -2\varepsilon & 0 & 0 \\
   0 & 0 & 0 & 0 & 0
   \end{array} \right] , \hspace{1cm}
 {\cal J}_2 =  \frac{1}{\hbar} \left[ \begin{array}{cccc}
   0 & -\varepsilon & 0 & -J \\
   \varepsilon & 0 & J & 0 \\
   0 & -J & 0 & \varepsilon \\
   J & 0 & -\varepsilon & 0
   \end{array} \right] .
\end{eqnarray}

The matrices $ {\cal G}_1(t) $ and $ {\cal G}_2(t) $ stemm
from the exciton-phonon interaction:
\begin{eqnarray}         % A2
 {\cal G}_1(t) =  \left[ \begin{array}{ccccc}
   0 & 0 & 0 & 0 & 0 \\
   0 & 0 & 0 & 0 & 0 \\
   A & C & E & -F & 0 \\
   B & -D & F & E & 0 \\
   0 & 0 & 0 & 0 & 0
   \end{array} \right] , \hspace{1cm}
 {\cal G}_2(t) =  \left[ \begin{array}{cccc}
   A_1 & -B_1 & C_1 & -D_1 \\
   B_1 & A_1 & D_1 & C_1 \\
   C_2 & -D_2 & A_2 & -B_2 \\
   D_2 & C_2 & B_2 & A_2
   \end{array} \right] .
\end{eqnarray}

The time-dependent coefficients $ A(t), \ldots, D_2(t) $
are given as follows
\begin{eqnarray}             % A3-A4
 A(t) &=& - \frac{J\varepsilon}{\Delta^2} G^2 \bar{g}_{2,1} +
  \frac{J}{2\Delta} G^2 \bar{g}_{3,2} , \nonumber \\
 B(t) &=& - \frac{J\varepsilon}{\Delta^2} G^2 \bar{g}_{2,2} -
  \frac{J}{2\Delta} G^2 \bar{g}_{3,1} , \nonumber \\
 C(t) &=& \frac{J\varepsilon}{\Delta^2} G^2 \bar{g}_{2,1} +
  \frac{J}{2\Delta} G^2 \bar{g}_{3,2} , \nonumber \\
 D(t) &=& \frac{J\varepsilon}{\Delta^2} G^2 \bar{g}_{2,2} -
  \frac{J}{2\Delta} G^2 \bar{g}_{3,1} , \nonumber \\
 E(t) &=& G^2 \bar{g}_{1,1} - 2\frac{J^2}{\Delta^2} G^2 \bar{g}_{2,1} ,
  \nonumber \\
 F(t) &=& -G^2 \bar{g}_{1,3} , \label{A1} \\
 A_1(t) &=& G^2 \bar{g}_{1,4} + \frac{J^2}{\Delta^2} G^2 \left[
  -\bar{g}_{2,4} + \bar{g}_{2,5} + \bar{g}_{2,10} \right] \nonumber , \\
 B_1(t) &=& G^2 \bar{g}_{1,8} + \frac{J^2}{\Delta^2} G^2 \left[ -\bar{g}_{2,6}
  - \bar{g}_{2,8} + \bar{g}_{2,9} \right] \nonumber , \\
 A_2(t) &=& G^2 \bar{g}_{1,7} + \frac{J^2}{\Delta^2} G^2 \left[ \bar{g}_{2,5}
  - \bar{g}_{2,7} - \bar{g}_{2,10} \right] \nonumber , \\
 B_2(t) &=& G^2 \bar{g}_{1,11} + \frac{J^2}{\Delta^2} G^2 \left[ \bar{g}_{2,6}
  + \bar{g}_{2,9} - \bar{g}_{2,11} \right] \nonumber , \\
 C_1(t) &=& \frac{J\varepsilon}{\Delta^2} G^2 \left[ \bar{g}_{2,4}
  - \bar{g}_{2,5}
  - \bar{g}_{2,10} \right] + \frac{J}{2\Delta} G^2 \left[ \bar{g}_{3,6} +
  \bar{g}_{3,8} - \bar{g}_{3,9} \right] , \nonumber \\
 D_1(t) &=& \frac{J\varepsilon}{\Delta^2} G^2 \left[ \bar{g}_{2,6} +
  \bar{g}_{2,8} -
  \bar{g}_{2,9} \right] + \frac{J}{2\Delta} G^2 \left[ -\bar{g}_{3,4} +
  \bar{g}_{3,5} + \bar{g}_{3,10}  \right] , \nonumber \\
 C_2(t) &=& \frac{J\varepsilon}{\Delta^2} G^2 \left[ \bar{g}_{2,5} -
  \bar{g}_{2,7} - \bar{g}_{2,10} \right] + \frac{J}{2\Delta} G^2 \left[
  -\bar{g}_{3,6} - \bar{g}_{3,9} + \bar{g}_{3,11} \right] , \nonumber \\
 D_2(t) &=& \frac{J\varepsilon}{\Delta^2} G^2 \left[ \bar{g}_{2,6} +
  \bar{g}_{2,9} - \bar{g}_{2,11} \right] + \frac{J}{2\Delta} G^2 \left[
  \bar{g}_{3,5} - \bar{g}_{3,7} - \bar{g}_{3,10}  \right] .
  \label{A2}
\end{eqnarray}

The functions
\begin{eqnarray}             % A5
 \bar{g}_{1,j}(t) &=& \int^{t-t_0}_0 d\tau \, g_j(\tau) , \nonumber \\
 \bar{g}_{2,j}(t,\Delta') &=& \int^{t-t_0}_0 d\tau \, g_j(\tau)\sin^2( \Delta'
  \tau ) , \nonumber \\
 \bar{g}_{3,j}(t,\Delta') &=& \int^{t-t_0}_0 d\tau \, g_j(\tau)\sin( 2\Delta'
  \tau ) , \hspace{1cm} j=1,\ldots,11
 \label{A3}
\end{eqnarray}
describe the response of the exciton subsystem to the phonon one.
The phonon subsystem is characterized by the functions
\footnote{In numerical calculations,
we assume that $ \hbar\Omega_k
G^i_k = G_i $ in one half of the $ k $--space and $ \hbar\Omega_k
G^i_{k} = G^{*}_{i} $ in the remaining half of the $ k $--space
($ i=1,2 $) (for details, see, Ref. \cite{diser}).
This assumption is in agreement with hermiticity of
$ \hat{H}_{\rm e-ph} $. Further, the mean numbers of
phonons $ n_{\rm B}(\hbar\Omega_k) $ are assumed to be
$ k $--independent ($ n_{\rm B}(\hbar\Omega_{k_0}) = n_{\rm B} $).
The remaining
summations $ \frac{1}{N} \sum_{k} \sin(\Omega_k\tau) $ and
$ \frac{1}{N} \sum_{k} \cos(\Omega_k\tau) $ in Eqs. (A6)
are replaced
by the expressions $ \sin(\Omega_{ph}\tau)\exp(-\gamma_{ph}\tau) $ and
$ \cos(\Omega_{ph}\tau)\exp(-\gamma_{ph}\tau) $, respectively.
The frequency
$ \Omega_{ph} $ then characterizes a mean phonon oscillation frequency
and
$ \gamma_{ph} $ describes damping originating in dephasing.}
\begin{eqnarray}             % A6
 g_1(\tau) &=& \frac{1}{G^2 N} \sum_k \Omega^2_k |G^1_k-G^2_k|^2
  \left[ 2 n_{\rm B}(\hbar\Omega_k) +1 \right] \cos(\Omega_k\tau) ,
  \nonumber \\
 g_2(\tau) &=& \frac{1}{G^2 N} \sum_k \Omega^2_k |G^1_k-G^2_k|^2
  \sin(\Omega_k\tau) , \nonumber \\
 g_3(\tau) &=& \frac{1}{G^2 N} \sum_k \Omega^2_k (G^1_k-G^2_k) (G^1_{-k}
  + G^2_{-k}) \sin(\Omega_k\tau) , \nonumber \\
 g_4(\tau) &=& \frac{1}{G^2 N} \sum_k \Omega^2_k |G^1_k|^2
  \left[ 2 n_{\rm B}(\hbar\Omega_k) +1 \right] \cos(\Omega_k\tau) ,
  \nonumber \\
 g_5(\tau) &=& \frac{1}{G^2 N} \sum_k \Omega^2_k {\rm Re} [G^1_k G^2_{-k}]
  \left[ 2 n_{\rm B}(\hbar\Omega_k) +1 \right] \cos(\Omega_k\tau) ,
  \nonumber \\
 g_6(\tau) &=& \frac{1}{G^2 N} \sum_k \Omega^2_k {\rm Im} [G^1_k G^2_{-k}]
  \left[ 2 n_{\rm B}(\hbar\Omega_k) +1 \right] \cos(\Omega_k\tau) ,
  \nonumber \\
 g_7(\tau) &=& \frac{1}{G^2 N} \sum_k \Omega^2_k |G^2_k|^2
  \left[ 2 n_{\rm B}(\hbar\Omega_k) +1 \right] \cos(\Omega_k\tau) ,
  \nonumber \\
 g_8(\tau) &=& \frac{1}{G^2 N} \sum_k \Omega^2_k |G^1_k|^2
  \sin(\Omega_k\tau) , \nonumber \\
 g_9(\tau) &=& \frac{1}{G^2 N} \sum_k \Omega^2_k {\rm Re} [G^1_k G^2_{-k}]
  \sin(\Omega_k\tau) , \nonumber \\
 g_{10}(\tau) &=& \frac{1}{G^2 N} \sum_k \Omega^2_k {\rm Im} [G^1_k G^2_{-k}]
  \sin(\Omega_k\tau) , \nonumber \\
 g_{11}(\tau) &=& \frac{1}{G^2 N} \sum_k \Omega^2_k |G^2_k|^2
  \sin(\Omega_k\tau) \label{A5} .
\end{eqnarray}

The constant $ G $ having the meaning of the
mean exciton--phonon interaction constant has been
introduced into Eqs. (A3) and (A4)
as well as into the definitions given in Eq. (A6)
in order to get $ g_1(\tau),
\ldots, g_{11}(\tau) $ dependent only on the dispersion of coupling
constants. The symbols $ {\rm Re} $ and $ {\rm Im} $
denote real and imaginary parts. The new symbol
$ \Delta' = \Delta /\hbar $ has been introduced here.

The time-dependent coefficients $ B(t) $ and $ D(t) $
given in Eq. (A3)
renormalize the transfer integral $ J $.
The coefficients $ A(t) $ and $ C(t) $ are important for
relaxation to equilibrium state (see, e.g.,
Refs. \cite{Cap1,Cap1a,Cap2}).

The matrices $ {\cal F}_1(t) $, $ {\cal F}_2(t) $,
$ {\cal F}_3(t) $, and $ {\cal F}_4(t) $ originate
in the interaction with an optical field:
\begin{eqnarray}         % A7
 {\cal F}_1(t) &=&  \left[ \begin{array}{ccccc}
   2\bar{M}_1 & 0 & 2\bar{O}_1 & 2\bar{O}_2 & -2M_1 \\
   0 & 2\bar{N}_1 & 2\bar{P}_1 & -2\bar{P}_2 & -2N_1 \\
   \bar{P}_1 & \bar{O}_1 & \bar{M}_1+\bar{N}_1 &
    -\bar{M}_2+\bar{N}_2 & -O_1-P_1 \\
   -\bar{P}_2 & \bar{O}_2 & \bar{M}_2-\bar{N}_2 &
    \bar{M}_1+\bar{N}_1 & P_2-O_2 \\
   -2\bar{M}_1 & -2\bar{N}_1 & -2\bar{P}_1-2\bar{O}_1 &
    2\bar{P}_2-2\bar{O}_2 & 2M_1+2N_1
   \end{array} \right] , \nonumber \\
 {\cal F}_2(t) &=&  \left[ \begin{array}{cccc}
   2K_1 & -2K_2 & 0 & 0 \\
   0 & 0 & 2L_1 & -2L_2 \\
   L_1 & -L_2 & K_1 & -K_2 \\
   -L_2 & -L_1 & K_2 & K_1 \\
   -2K_1 & 2K_2 & -2L_1 & 2L_2
   \end{array} \right] , \nonumber \\
 {\cal F}_3(t) &=&  \left[ \begin{array}{ccccc}
   -K_1 & 0 & -L_1 & L_2 & K_1 \\
   K_2 & 0 & L_2 & L_1 & -K_2 \\
   0 & -L_1 & -K_1 & -K_2 & L_1 \\
   0 & L_2 & K_2 & -K_1 & -L_2
   \end{array} \right] , \nonumber \\
 {\cal F}_4(t) &=&  \left[ \begin{array}{cccc}
   2M_1+N_1-2\tilde{M}_1 & 2M_2+N_2-2\tilde{M}_2 &
    O_1-\tilde{O}_1-\tilde{P}_1 & O_2-\tilde{O}_2-\tilde{P}_2 \\
   -2M_2-N_2-2\tilde{M}_2 & 2M_1+N_1+2\tilde{M}_1 &
    -O_2-\tilde{O}_2-\tilde{P}_2 & O_1+\tilde{O}_1+\tilde{P}_1 \\
   P_1-\tilde{O}_1-\tilde{P}_1 & P_2-\tilde{O}_2-\tilde{P}_2 &
    2N_1+M_1-2\tilde{N}_1 & 2N_2+M_2-2\tilde{N}_2 \\
   -P_2-\tilde{O}_2-\tilde{P}_2 & P_1+\tilde{O}_1+\tilde{P}_1 &
    -2N_2-M_2-2\tilde{N}_2 & 2N_1+M_1+2\tilde{N}_1
   \end{array} \right] . \nonumber \\
    & &
\end{eqnarray}

The coefficients $ K_1(t) $, $ K_2(t) $, $ L_1(t) $,
and $ L_2(t) $ describing the influence
of the coherent part of an optical field have the form
(the constants $ \tilde{F}^1_{K_0} $ and $ \tilde{F}^2_{K_0} $
are assumed to be real):
\begin{eqnarray}             % A8
 K_1(t) &=& -\omega_{K_0} \tilde{F}^1_{K_0} {\rm Im} \left[ \tilde{\cal A}(t)
  \exp (i\delta't) \right] , \nonumber \\
 K_2(t) &=& \omega_{K_0} \tilde{F}^1_{K_0} {\rm Re} \left[ \tilde{\cal A}(t)
  \exp (i\delta't) \right] , \nonumber \\
 L_1(t) &=& -\omega_{K_0} \tilde{F}^2_{K_0} {\rm Im} \left[ \tilde{\cal A}(t)
  \exp (i\delta't) \right] , \nonumber \\
 L_2(t) &=& \omega_{K_0} \tilde{F}^2_{K_0} {\rm Re} \left[ \tilde{\cal A}(t)
  \exp (i\delta't) \right] .
 \label{A6}
\end{eqnarray}
The symbol $ \delta' $ denotes the frequency mismatch
($ \delta' = (E+\varepsilon)/\hbar -\omega_{K_0} $) and
the envelope $ \tilde{\cal A}(t) $ of the field is defined as
follows,
\begin{equation}        % A9
 \tilde{\cal A}(t) = {\cal A}(t) \exp(i\omega_{K_0}t) .
\end{equation}

The coefficients $ M_1(t), \ldots, P_2(t) $ reflect
statistical properties of the optical field (noise)
and can be expressed in the form
\begin{eqnarray}             % A10
 M_1(t) &=& (\tilde{F}^1_{K_0})^2  i_1 - (\tilde{F}^1_{K_0})^2
  \frac{\varepsilon}{\Delta}  i_4 - \tilde{F}^1_{K_0} \tilde{F}^2_{K_0}
  \frac{J}{\Delta} i_4 , \nonumber \\
 M_2(t) &=& (\tilde{F}^1_{K_0})^2  i_3 + (\tilde{F}^1_{K_0})^2
  \frac{\varepsilon}{\Delta}  i_2 + \tilde{F}^1_{K_0} \tilde{F}^2_{K_0}
  \frac{J}{\Delta} i_2 , \nonumber \\
 N_1(t) &=& (\tilde{F}^2_{K_0})^2  i_1 + (\tilde{F}^2_{K_0})^2
  \frac{\varepsilon}{\Delta}  i_4 - \tilde{F}^1_{K_0} \tilde{F}^2_{K_0}
  \frac{J}{\Delta} i_4 , \nonumber \\
 N_2(t) &=& (\tilde{F}^2_{K_0})^2  i_3 - (\tilde{F}^2_{K_0})^2
  \frac{\varepsilon}{\Delta}  i_2 + \tilde{F}^1_{K_0} \tilde{F}^2_{K_0}
  \frac{J}{\Delta} i_2 , \nonumber \\
 O_1(t) &=& - (\tilde{F}^1_{K_0})^2 \frac{J}{\Delta} i_4 + \tilde{F}^1_{K_0}
  \tilde{F}^2_{K_0}  i_1 + \tilde{F}^1_{K_0} \tilde{F}^2_{K_0}
  \frac{\varepsilon}{\Delta} i_4 , \nonumber \\
 O_2(t) &=& (\tilde{F}^1_{K_0})^2 \frac{J}{\Delta} i_2 + \tilde{F}^1_{K_0}
  \tilde{F}^2_{K_0}  i_3 - \tilde{F}^1_{K_0} \tilde{F}^2_{K_0}
  \frac{\varepsilon}{\Delta} i_2 , \nonumber \\
 P_1(t) &=& - (\tilde{F}^2_{K_0})^2 \frac{J}{\Delta} i_4 + \tilde{F}^1_{K_0}
  \tilde{F}^2_{K_0}  i_1 - \tilde{F}^1_{K_0} \tilde{F}^2_{K_0}
  \frac{\varepsilon}{\Delta} i_4 , \nonumber \\
 P_2(t) &=& (\tilde{F}^2_{K_0})^2 \frac{J}{\Delta} i_2 + \tilde{F}^1_{K_0}
  \tilde{F}^2_{K_0}  i_3 + \tilde{F}^1_{K_0} \tilde{F}^2_{K_0}
  \frac{\varepsilon}{\Delta} i_2 .
 \label{A8}
\end{eqnarray}

The functions $ i_1(t,\Delta',\delta'), \ldots, i_4(t,\Delta',\delta') $
characterize the response of the exciton subsystem to the photon
field:
\begin{eqnarray}             % A11
 i_1(t,\Delta',\delta') &=& \omega^2_{K_0} \int^t_{t_0} d\tau \,
  \cos\left[\Delta'(t-\tau)\right] {\rm Re} \left[ \delta\tilde{N}(t,\tau)
  \exp\left[i\delta'(t-\tau)\right] \right] , \nonumber \\
 i_2(t,\Delta',\delta') &=& \omega^2_{K_0} \int^t_{t_0} d\tau \,
  \sin\left[\Delta'(t-\tau)\right] {\rm Re} \left[ \delta\tilde{N}(t,\tau)
  \exp\left[i\delta'(t-\tau)\right] \right] , \nonumber \\
 i_3(t,\Delta',\delta') &=& \omega^2_{K_0} \int^t_{t_0} d\tau \,
  \cos\left[\Delta'(t-\tau)\right] {\rm Im} \left[ \delta\tilde{N}(t,\tau)
  \exp\left[i\delta'(t-\tau)\right] \right] , \nonumber \\
 i_4(t,\Delta',\delta') &=& \omega^2_{K_0} \int^t_{t_0} d\tau \,
  \sin\left[\Delta'(t-\tau)\right] {\rm Im} \left[ \delta\tilde{N}(t,\tau)
  \exp\left[i\delta'(t-\tau)\right] \right].
  \label{A9}
\end{eqnarray}
The photon field correlation function $ \delta\tilde{N}(t,\tau) $
is of the form:
\begin{equation}             % A12
 \delta\tilde{N}(t,\tau) = \delta N(t,\tau) \exp\left[ i\omega_{K_0}(t-\tau)
  \right] .
  \label{A10}
\end{equation}

The coefficients $ \bar{M}_1(t), \ldots, \bar{P}_2(t) $  are
defined similarly as the coefficients
$ M_1(t), \ldots, P_2(t) $ in Eqs. (A10) and (A11);
only the photon field correlation function
\begin{equation}             % A13
 \delta\tilde{N}_v(t,\tau) = \delta N_v(t,\tau) \exp\left[
  i\omega_{K_0}(t-\tau) \right]
 \label{A11}
\end{equation}
occurs in Eq. (A11) instead of $ \delta\tilde{N}(t,\tau) $.
Thus, the coefficients with bars include in addition
effects of vacuum fluctuations.

Also the coefficients $ \tilde{M}_1(t), \ldots, \tilde{P}_2(t) $
are defined similarly as the coefficients $ M_1(t), \ldots, P_2(t) $
in Eqs. (A10) and (A11): only the expression
$ \delta\tilde{N}(t,\tau) \exp\left[i\delta'(t-\tau)\right] $
in Eq. (A11) must be replaced by the expression $ \delta\tilde{N}_a(t,\tau)
\exp\left[-i\delta'(t+\tau)\right] $, where
\begin{equation}             % A14
 \delta\tilde{N}_a(t,\tau) = \delta N_a(t,\tau) \exp\left[ -i\omega_{K_0}
  (t+\tau) \right] .
 \label{A12}
\end{equation}

The coefficients $ \bar{O}_2(t) $ and $ \bar{P}_2(t) $ renormalize
the transfer integral $ J $ both
at the positions (4,1), (1,4) and (4,2), (2,4) in the matrix
$ {\cal J}_1 $,
in contrast to the coefficients originating in the exciton--phonon
interaction.

The influence of the coefficient $ \tilde{M}_1(t) $ ($ \tilde{N}_1(t) $)
in the equations
for $ \rho_{1r}(t) $ and $ \rho_{1i}(t) $ ($ \rho_{2r}(t) $ and
$ \rho_{2i}(t) $) is remarkable.
When, e.g., $ \tilde{M}_1(t) $ ($ \tilde{N}_1(t) $) is negative, it
represents damping of $ \rho_{1r}(t) $ ($ \rho_{2r}(t) $), but at the
same time amplification of $ \rho_{1i}(t) $
($ \rho_{2i}(t) $).
This property is connected with phase relations in the photon field
reflected by $ \delta \tilde{N}_a(t,\tau) $ given in Eq. (A14).

\end{document}